\documentclass[preprint]{revtex4-2}
\usepackage{amsmath,amssymb,amsfonts}
\usepackage{graphicx}
\usepackage{geometry}
\usepackage{fancyhdr}
\usepackage{citesort}
\usepackage{xcolor}
\usepackage{soul}
\usepackage{lineno}
\AtBeginDocument{\thispagestyle{fancy}}  % this is needed for front page
\begin{document}
\title{Structural origin of relaxation in dense colloidal suspensions}

\author{Ratimanasee Sahu$^{1}$, Mohit Sharma$^{2}$,$^4$ , Peter Schall$^{3}$, Sarika Maitra Bhattacharyya$^{2}$,$^4$\textsuperscript{*}, Vijayakumar Chikkadi$^{1}$\textsuperscript{$\dagger$}}

\affiliation{
$^1$ Physics Division, Indian Institute of Science Education and Research Pune, Pune-411008, India.\\
$^2$ Polymer Science and Engineering Division, CSIR-National Chemical Laboratory, Pune-411008, India.\\
$^3$ Institute of Physics, University of Amsterdam, 1098 XH Amsterdam, The Netherlands.\\
$^4$ Academy of Scientific and Innovative Research (AcSIR), Ghaziabad 201002, India\\
}

\begin{abstract}
Amorphous solids relax via slow molecular rearrangement induced by thermal fluctuations or applied stress. Microscopic structural signatures predicting these structural relaxations have been long searched for but have so far only been found in dynamic quantities such as vibrational quasi-localized soft modes or with structurally trained neural networks. A physically meaningful structural quantity remains elusive. Here, we introduce a structural order parameter derived from the mean-field caging potential experienced by the particles due to their neighbors, which reliably predicts the occurrence of structural relaxations. The structural parameter, derived from density functional theory, provides a measure of susceptibility to particle rearrangements that can effectively identify weak or defect-like regions in disordered systems. Using experiments on dense colloidal suspensions, we demonstrate a strong correlation between this order parameter and the structural relaxations of the amorphous solid. In quiescent suspensions, this correlation increases with density, when particle rearrangements become rarer and more localized. In sheared suspensions, the order parameter reliably pinpoints shear transformations; the applied shear weakens the caging potential due to shear-induced structural distortions, causing the proliferation of plastic deformation at structurally weak regions. Our work paves the way to a structural understanding of the relaxation of a wide range of amorphous solids, from suspensions to metallic glasses.
\end{abstract}

\maketitle

%\textcolor{blue} {
%\section*{Significance Statement}
%Defects play a crucial role in determining the electrical and mechanical properties of various materials. Understanding their impact is essential for engineering materials with specific mechanical properties. Although the influence of defects on the plastic deformation of ordered crystals is well-established, it remains a significant challenge in amorphous solids. Defining a structural measure that can identify defect-like regions susceptible to failure under applied stress remains an outstanding problem. We propose a novel structural order parameter to identify such weak regions in dense suspensions and perform experiments to demonstrate its effectiveness in pinpointing regions where particles rearrange or plastic events occur due to applied shear or thermal fluctuations.}\\\\

\section*{Introduction}
The hallmark of amorphous solids is their slow structural relaxation, which occurs many orders of magnitude slower than the molecular relaxation time. This relaxation is dynamically heterogeneous \cite{Weitz00, Biroli11, Saarloos11, Garrahan11, Durian07}, related to the disordered structure of the glass, with dynamic time scales differing by orders of magnitude across the sample. A major effort has been to link this relaxation to structural hallmarks that define locally weak regions from which the relaxations originate. Unlike crystals, for which dislocations can be identified as topological defects that carry plastic relaxation within the long-range ordered lattice, no such structural measures exist for glasses, which possess only short-range order. Even when the amorphous solid is subject to external shear and the deformation is localized to regions referred to as shear transformed zones \cite{Langer98, Spaepen07}, no structural measure has been found to predict their occurrence. While the variation in relaxation dynamics is believed to originate from the diverse atomic environment of the disordered amorphous structure, with some regions being softer than others and thus more susceptible to failure under external stresses, no unique signature has yet been identified. Even though these localized shear transformations have been integral to the earliest models of glasses \cite{Cohen61, Grest79, Ronchetti83} and the plasticity of amorphous solids \cite{Argon79, Spaepen77, Langer98}, and several structural measures including free volume \cite{Cohen61,Grest79, Spaepen77}, elastic properties of the system \cite{Ma14, Falk16, Dyre98, Barrat09}, locally favored structures \cite{Royall13, Tanaka12, Royall18}, two-body excess entropy \cite{ Parrinello17_1, Parrinello17_2, Arratia20}, and soft modes of vibration \cite{Reichman08, Harrowell06, Bonn11, Liu11} were proposed to identify them, they have had limited success to identify structural relaxations under shear or thermal fluctuations \cite{Manning20,Chen16}. Although several other measures of plasticity were tested for their predictive efficacy \cite{Manning20}, only a small subset of these measures is applicable to experimental systems such as colloidal suspensions. \\

In recent years, novel methods based on machine learning techniques have been employed to investigate correlations between structure and particle dynamics. Liu and coworkers have made pioneering contributions using supervised learning techniques \cite{Liu15, Liu16, Liu16_2, Liu17, Liu18, Reichman20, Liu21, Liu21_1, Liu22, Liu23, Liu22_1}. They utilized support vector machines and a set of structural descriptors to define a "softness" parameter, which helps identify particles that are likely to rearrange. This was crucial in establishing a structural approach to understanding dynamic relaxation in glassy systems. These ideas were successfully tested not only in computer simulations but also in experiments \cite{Liu17,Yodh19,Ganapathy21}. Furthermore, these methods were extended to identify defects in amorphous solids \cite{Liu22}, understand atomistic motion in grain boundaries \cite{Liu18}, and predict the formation of shear transformation zones—considered plasticity carriers—in sheared amorphous solids using the softness parameter \cite{Liu21}. These concepts have also been applied to propose novel elasto-plastic models based on softness \cite{Liu21_1}. \\

These studies have inspired the development of other efficient methods, such as graph neural networks and physics-inspired deep neural network methods, to explore the physics of glasses \cite{Kohli20,Berthier23}. Recent research has used advanced machine learning techniques and SWAP Monte Carlo methods for efficient exploration of potential energy landscapes, providing evidence of quantum tunneling two-level systems in low-temperature glasses \cite{Zamponi23}. Additionally, these studies have led to the development of unsupervised techniques to understand the correlation between a structural order parameter and dynamical heterogeneity in amorphous systems \cite{Filion20, Filion21, Filion23}.\\

A few other order parameters have been successful in describing structural heterogeneity and correlating well with the dynamics in the systems but have not yet been explored for systems under applied shear. Tong and Tanaka have proposed an order parameter that is many body in nature and correlates well with the dynamics in 2D and 3D quiescent systems \cite{Tanaka18-1,Tanaka19,Tanaka20}. Based on rigorous dynamic density functional theory \cite{Wolynes87, Schweizer05, Brader12}, recently, some of us proposed a mean-field microscopic theory of softness \cite{Bhattacharyya17, Bhattacharyya21, Bhattacharyya22}. This formulation assumes each particle to be caged by its neighbours, as described by the structure of the liquid. The inverse depth of this mean-field caging potential served as a structural order parameter (SOP), and strong correlations between this SOP and the dynamics were found at low temperatures, where the system relaxed slowly. 
It was also shown that the SOP captures both enthalpic and entropic effects, the former playing a central role in the dynamics in attractive systems and the latter in repulsive systems, suggesting that this SOP is well equipped to be a good predictor of dynamics for both attractive and repulsive systems \cite{Bhattacharyya23}. However, the SOP was never tested in experiments nor for systems under applied shear, so its applicability and predictive power remain largely unexplored. This is notwithstanding the appeal of the order parameter to provide a simple, physically intuitive quantity directly obtained from the structure, conveniently implementable in experimental studies. \\

Here, we apply the structural order parameter to experiments on colloidal glasses and demonstrate that is an excellent descriptor to reliably pinpoint structural relaxations in quiescent and sheared conditions. We investigate both dense monolayers in quasi two-dimensional measurements and bulk suspensions in three dimensions, and employ dynamic density functional theory to compute the mean field caging potential and its inverse depth, which we define as the structural order parameter (SOP). The caging potential depth indicates the stiffness of the nearest-neighbor cage, and its inverse is the compliance of the nearest neighbor environment. Using this order parameter, we demonstrate a direct relation between the structure and dynamics of the systems. With increasing particle density, the caging potential depth increases, indicating a more defined structure, and leading to strong correlations between the SOP and the particle dynamics. We further apply the SOP to sheared suspensions and find that it reliably predicts the location of shear transformations. The applied shear lowers the caging potential and increases the SOP until shear transformations proliferate in regions of maximum SOP and the suspension yields. A quantification of these correlations confirms strong correlation between the structural order parameter and plastic deformation. The order parameter, therefore, provides a unique structural identifier of plastic regions in amorphous solids analogous to dislocations in crystals. 
%\com{Some more implications here ...?} \\

\begin{figure*}[t!]
\centering
\begin{tabular}{lll}
\includegraphics[width=.23\textwidth]{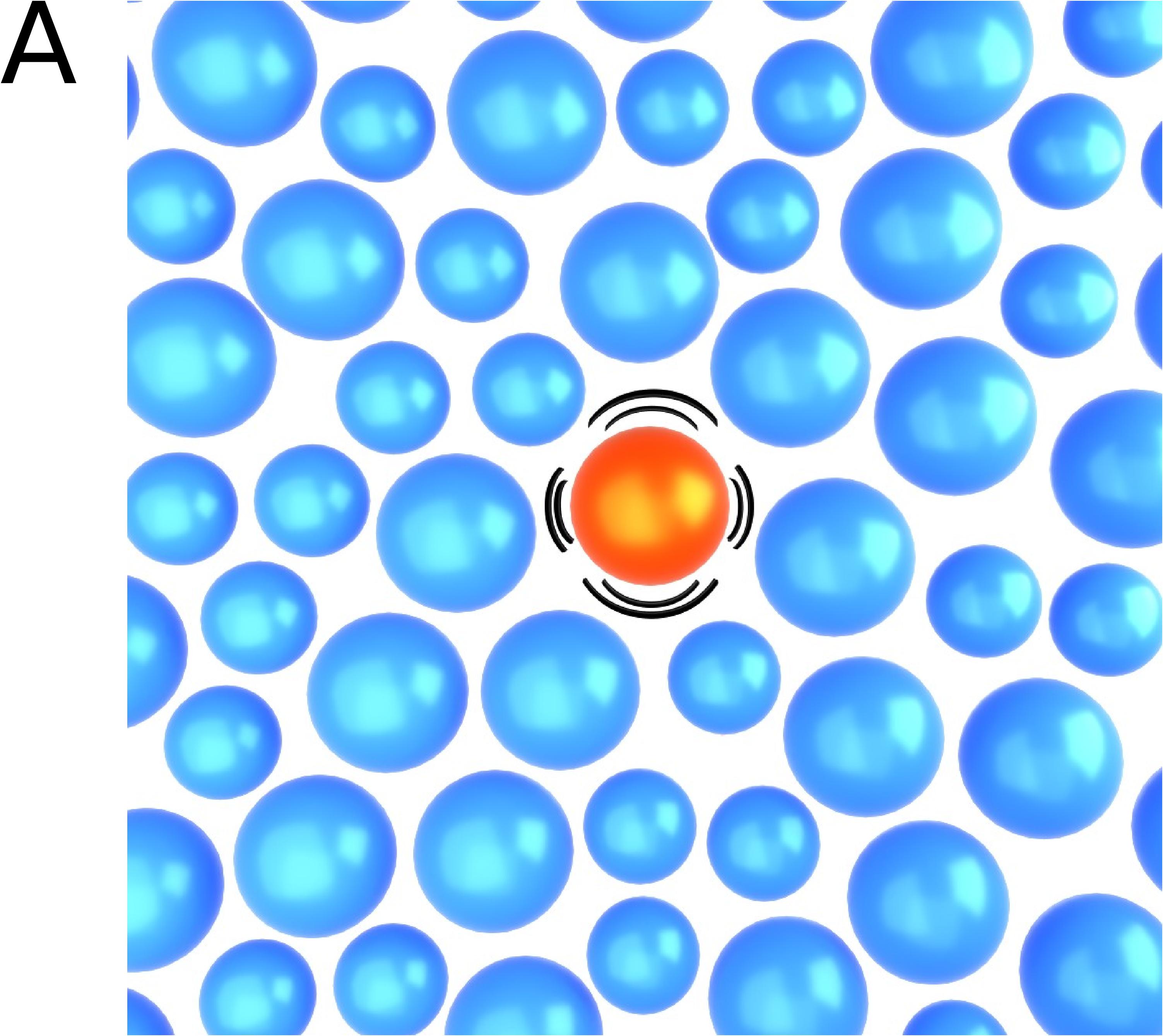}&
\includegraphics[width=.227\textwidth]{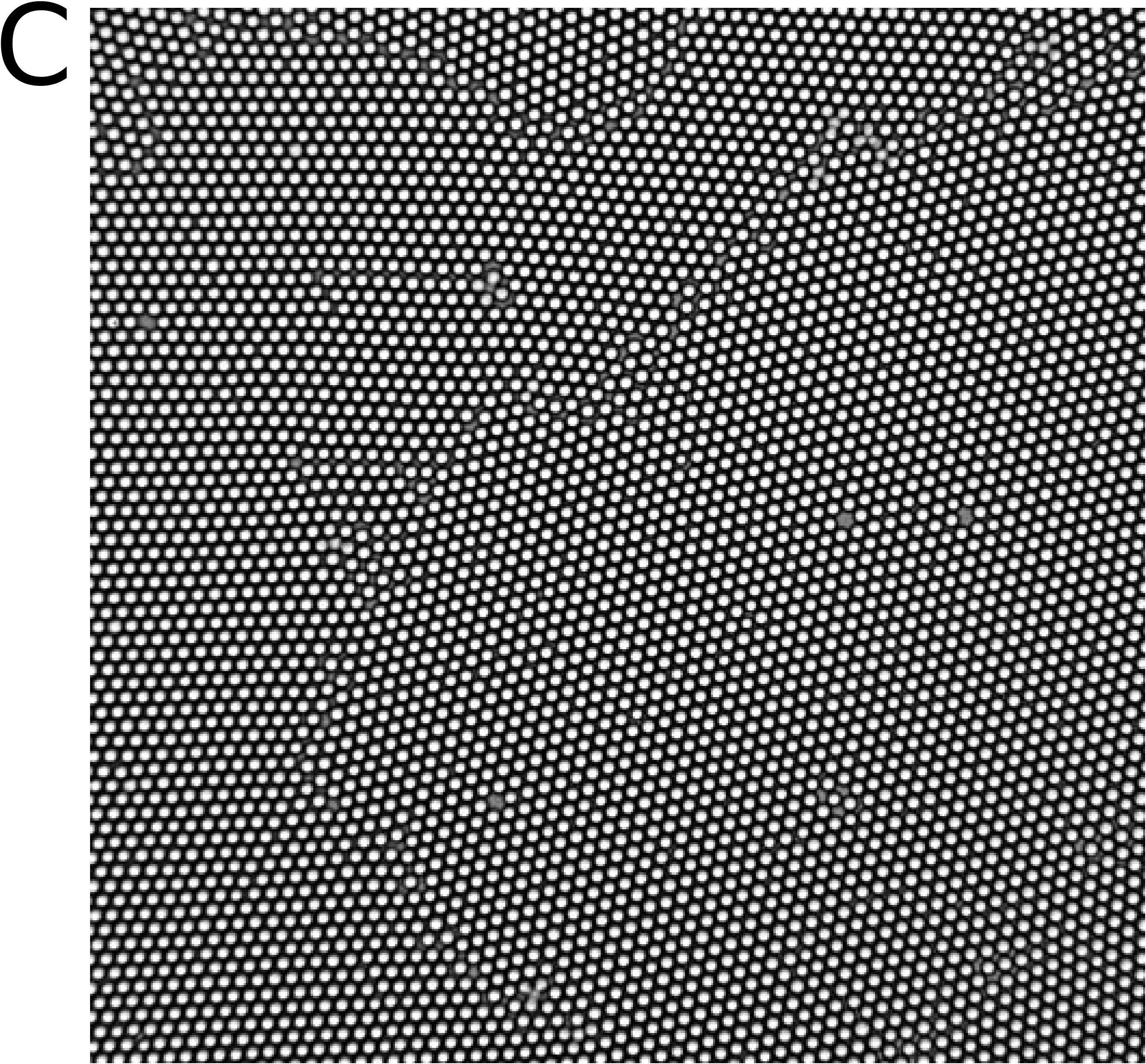}&
\includegraphics[width=.27\textwidth]{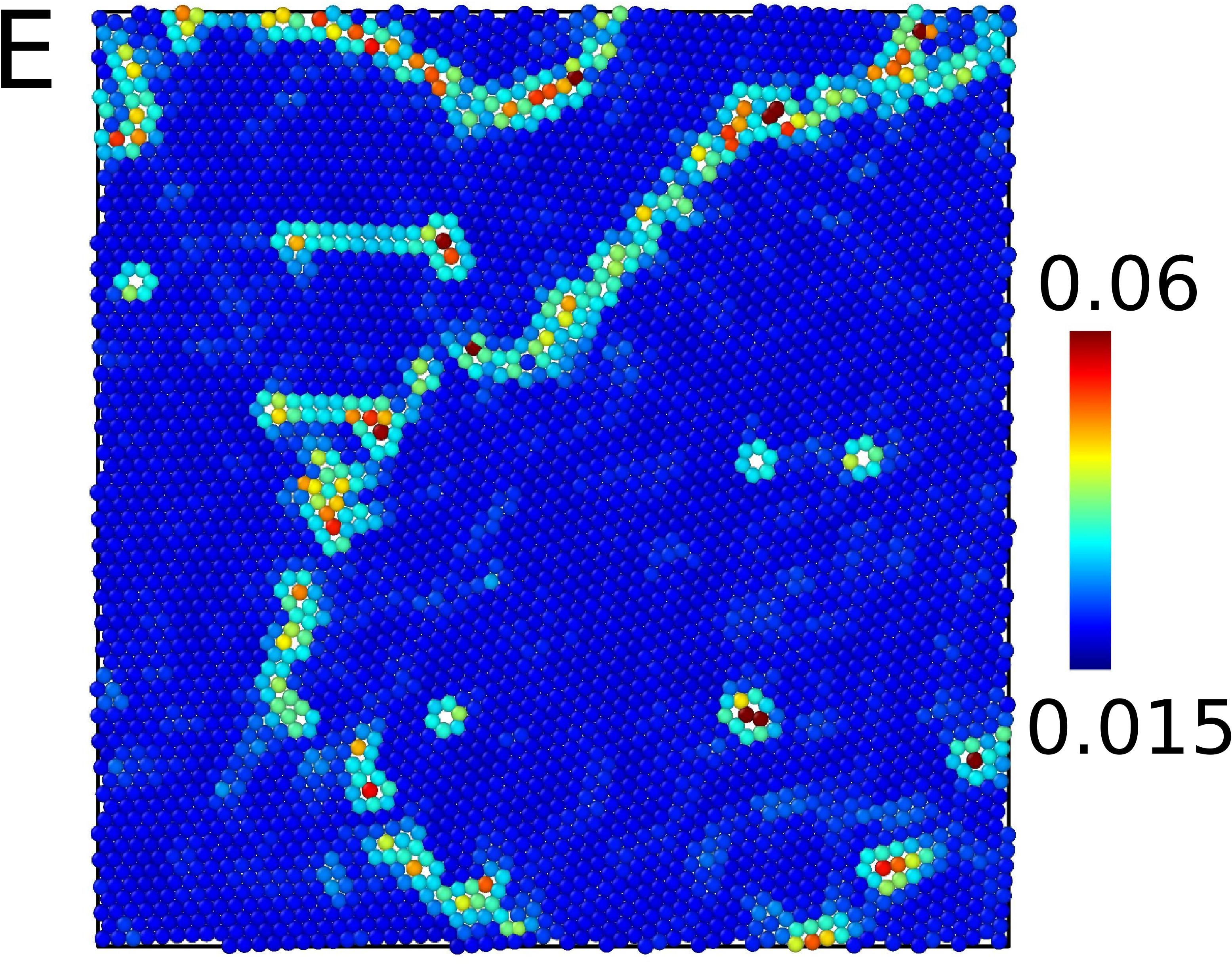}\\
\includegraphics[width=.26\textwidth]{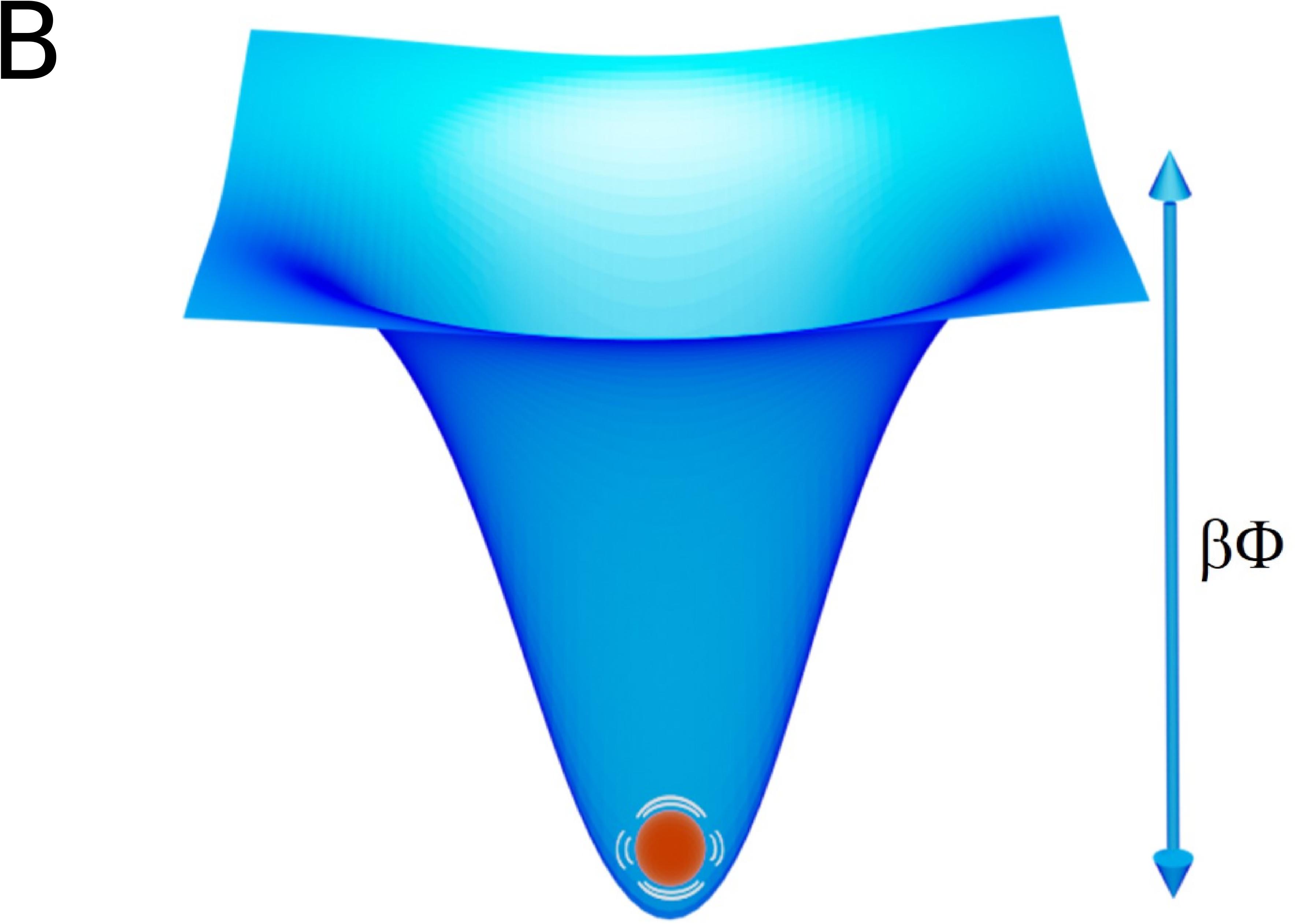}&
\includegraphics[width=.25\textwidth]{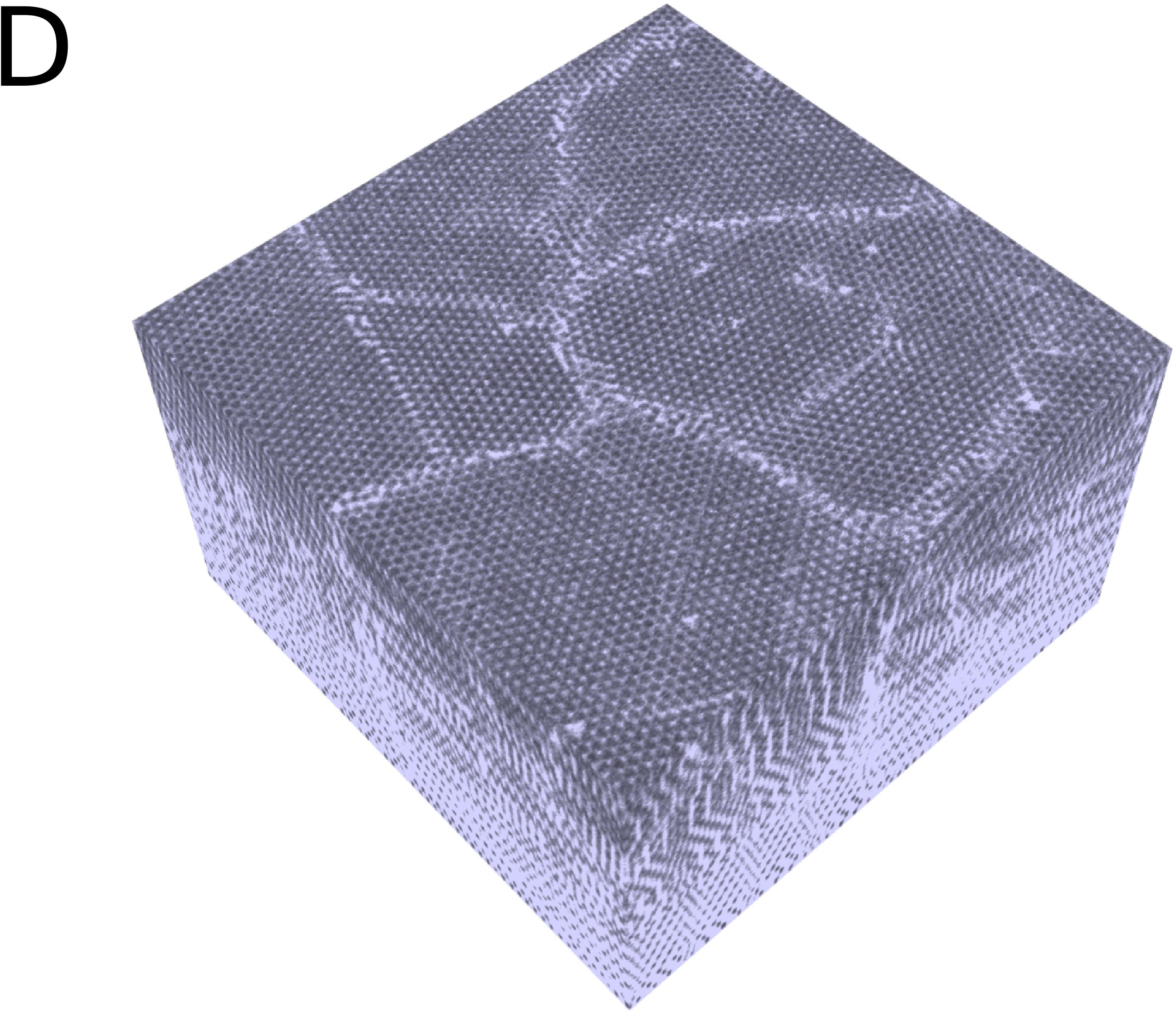}&
\includegraphics[width=.264\textwidth]{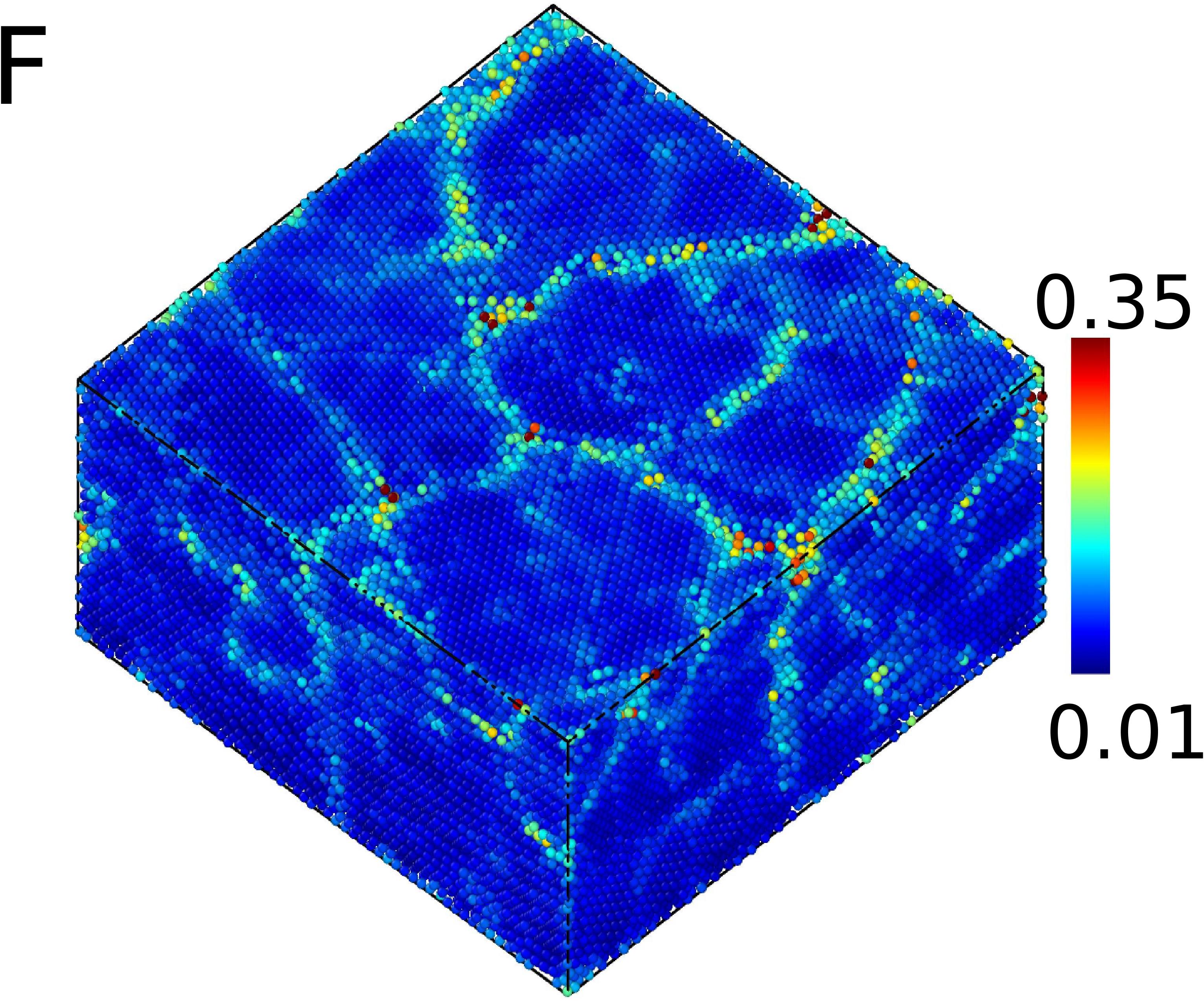}\\
\end{tabular}
\caption{ Mean-field caging potential and structural order parameter. (A) and (B) Schematics of a colloidal monolayer with a tagged particle (red), undergoing thermal fluctuations in a frozen background of the neighbouring particles (A), and the mean-field caging potential experienced by the tagged particle due to the frozen background of neighbouring particles (B). The scaled depth of the potential $\beta\Phi$ is obtained from the structure of the local neighborhood. (C) and (D) a Bright-field image of a mono-layer colloidal crystal and a confocal image of a 3D colloidal crystal in a fluorescent solvent, respectively. The false color in (D) highlights the grain boundaries and defects. The size of silica particles in the monolayer (C) is $3\mu m$, and the field of view is $280*280 \mu m $. The particles are $1\mu m$, and the field of view is $ 62*62*36 \mu m$ in (D). (E) and (F) The particles in the crystals are color coded based on the magnitude of their structural order parameter $S^i$. Blue color indicates a small SOP, and red color indicates a large SOP.}
\label{Fig1}
\end{figure*}

\section*{Caging-potential of dense colloidal suspensions}

A tagged particle in its neighbor environment is shown in Fig. \ref{Fig1}\emph{A}. According to the mean-field approximation, the caging potential of the particle is calculated assuming that the background is frozen while the particle is undergoing short-time dynamics \cite{Bhattacharyya17, Bhattacharyya21, Bhattacharyya22}. Using the Ramakrishnan-Yussouff free energy functional \cite{Yussouff79}, we define the mean-field caging potential \cite{Bhattacharyya21, Bhattacharyya17, Bhattacharyya22} felt by the particle due to the frozen background which now depends on the structure of the liquid. The details of the calculation of the mean field caging potential are described in earlier works\cite{Bhattacharyya21}, and a brief outline is presented in the supporting information. A cartoon of the potential experienced by the particle is shown in Fig. \ref{Fig1}\emph{B}. The form of the potential obtained in our experiments is shown in Fig.S1B in the supporting information.  \\

The absolute value of the depth of the caging potential felt by a particle in a system is given by \cite{Bhattacharyya22}, 
\begin{equation}\label{eq:1}
 \beta \Phi^{i}= \rho \int \mathbf{dr}~C^{i}(r) g^{i}(r),
\end{equation}
\noindent
where $\rho$ is the density, $g^{i}(r)$ is the particle level radial distribution function (RDF), and $C^{i}(r)$ is the direct correlation function which, via the hypernetted chain approximation \cite{Hansen}, is expressed in terms of the RDF as $C^{i}(r) \approx g^i(r)-1$ (see Section 1B and 1C in the SI for details). Note that this form of the direct correlation function allows us to use the formulation for systems where the interaction potential between the particles is unknown. Although the depth of the caging potential in Eq.1 is derived from microscopic density functional theory, it has a simple, intuitive meaning: The RDF provides information on the local arrangement of the particles around a tagged particle, and the direct correlation function is the effective short-range interaction potential between the tagged particle and its neighbours. Thus, the product of the two functions provides the caging potential the tagged particle feels due to its neighbours. For a more structured environment, $g(r)$ has a more pronounced first peak, and the potential will be deeper, while for a less structured environment, the first peak of $g(r)$ is less pronounced and the potential will be shallower. We exploit this relation between the local structure and the local depth of the potential to define the structural order parameter on a particle level $S^i \propto 1/\beta \Phi^i$, as a measure of the local stiffness based on the particle's immediate neighborhood.\\

As a proof of concept, we first apply this order parameter to colloidal crystals. We use aqueous suspensions of silica particles to prepare monolayer and bulk colloidal polycrystals. Microscope images are shown in Figs.\ref{Fig1}\emph{C} and \ref{Fig1}\emph{D}. The size of the particles in 2D crystals is $3\mu m$, and it is $1 \mu m$ in 3D crystals. The area fraction in 2D is $0.74$, and the volume fraction in 3D is $0.59$, see Materials and Methods for details. We compute the SOP of the particles from the structure of the local neighborhood using Eq.1. A map of the resulting structural order parameter is shown in Figs.\ref{Fig1}\emph{E} and \ref{Fig1}\emph{F}, where particle colour indicates its SOP value. Clearly, the order parameter can identify particles at the grain boundaries from their high SOP values: particles at the grain boundaries have a less defined environment and wider cages, and correspondingly a higher SOP. The same is true for other defects - line and point defects - in the 2D crystal, which is well-identified. Note that the order parameter is purely structural in nature; no input is needed from the dynamics of the system to describe it. \\
%Unlike the bond orientation order parameter that provides geometric information, our SOP is a measure of the local compliance.\\

\begin{figure*}
\centering
\begin{tabular}{lll}
\includegraphics[width=.33\textwidth]{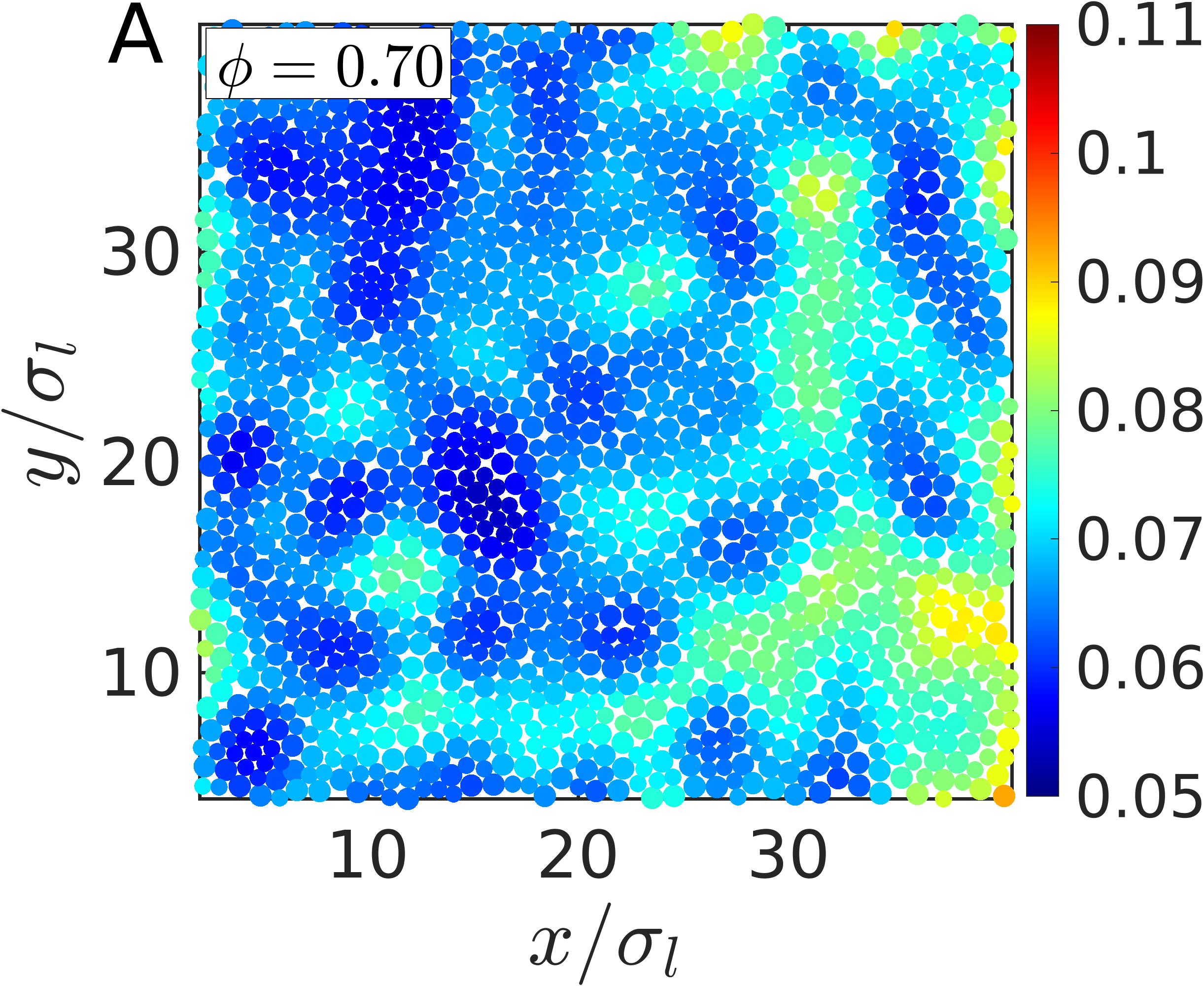}&
\includegraphics[width=.3\textwidth]{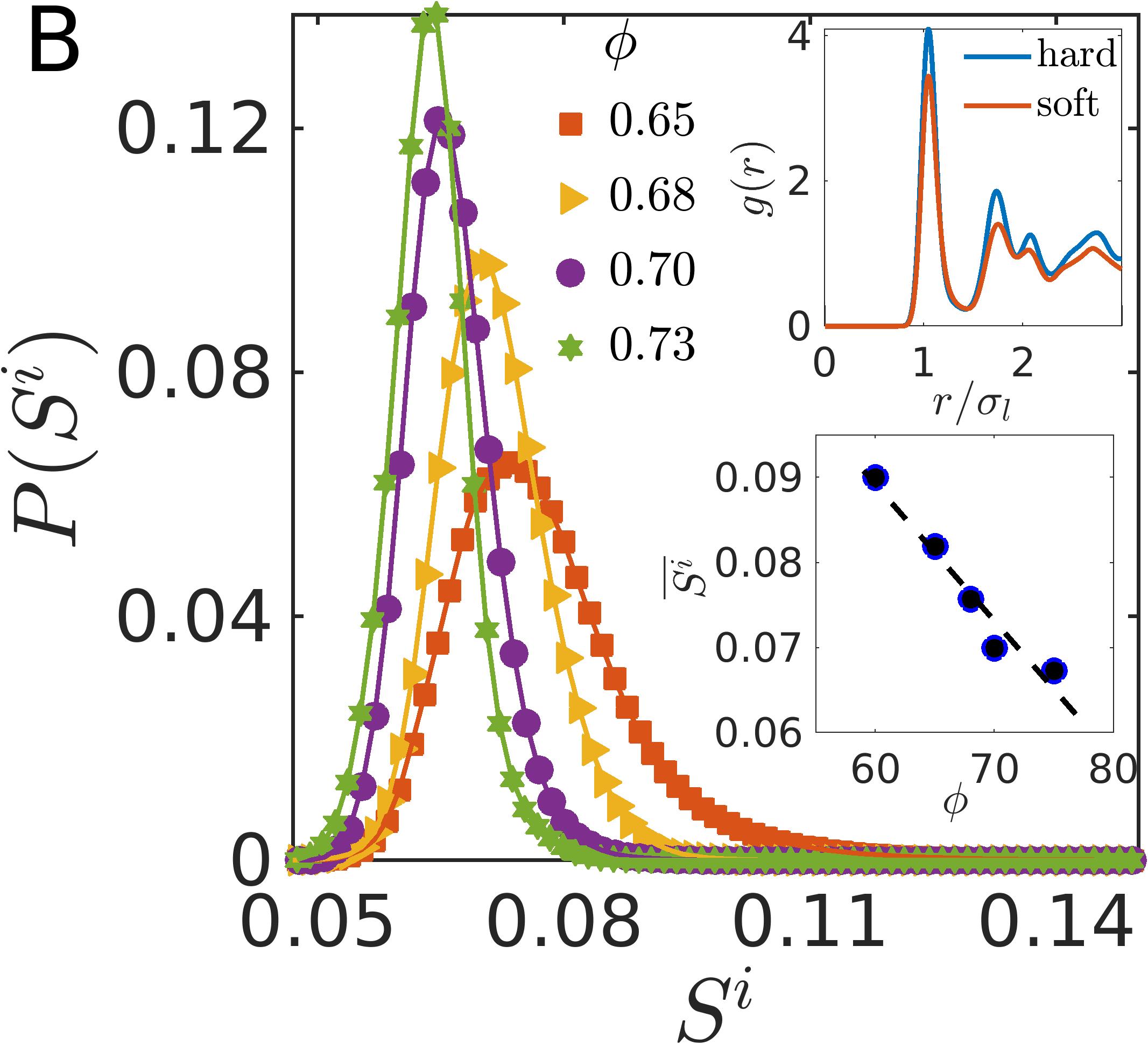}&
\includegraphics[width=.29\textwidth]{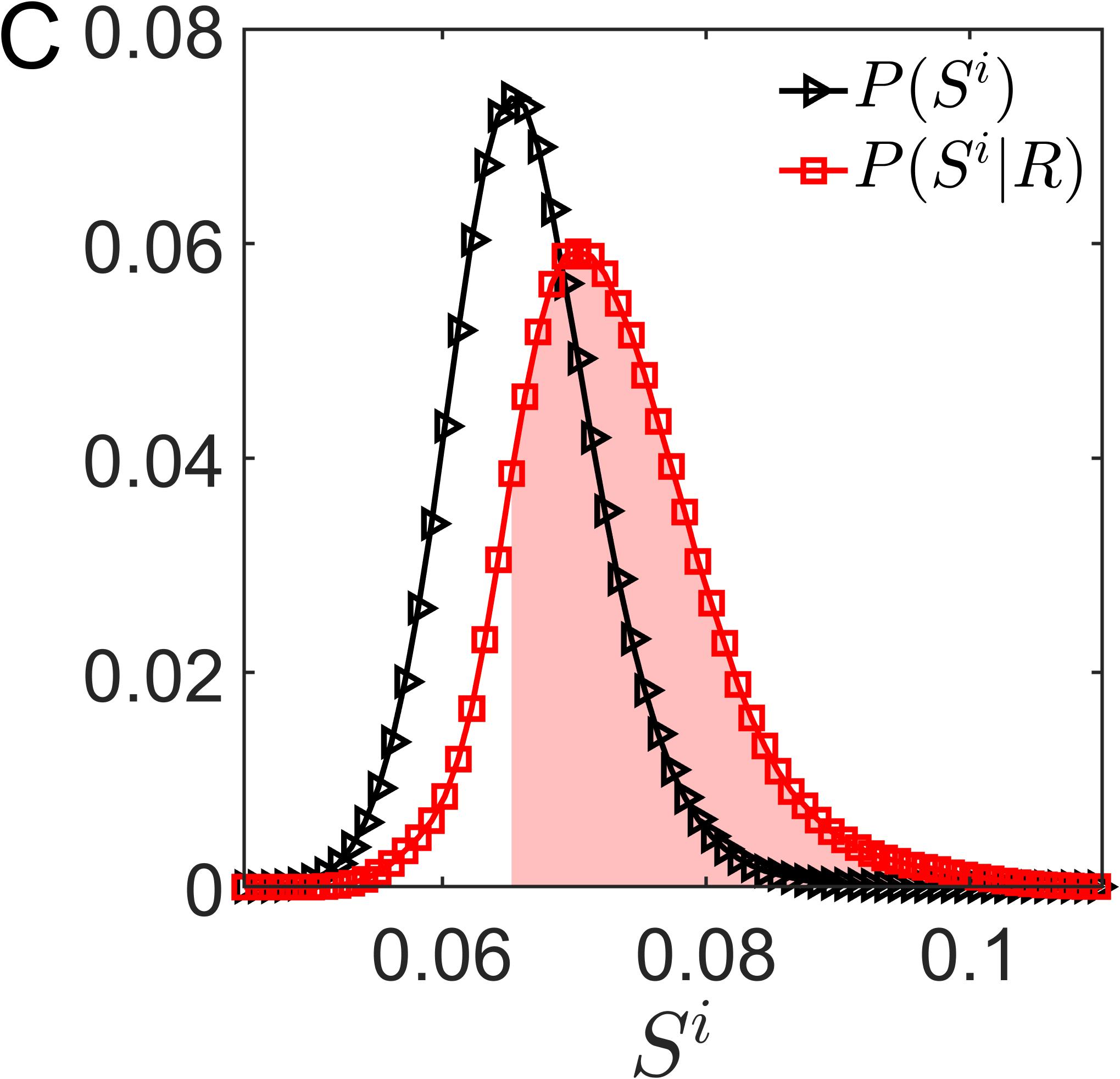}\\
\includegraphics[width=.335\textwidth]{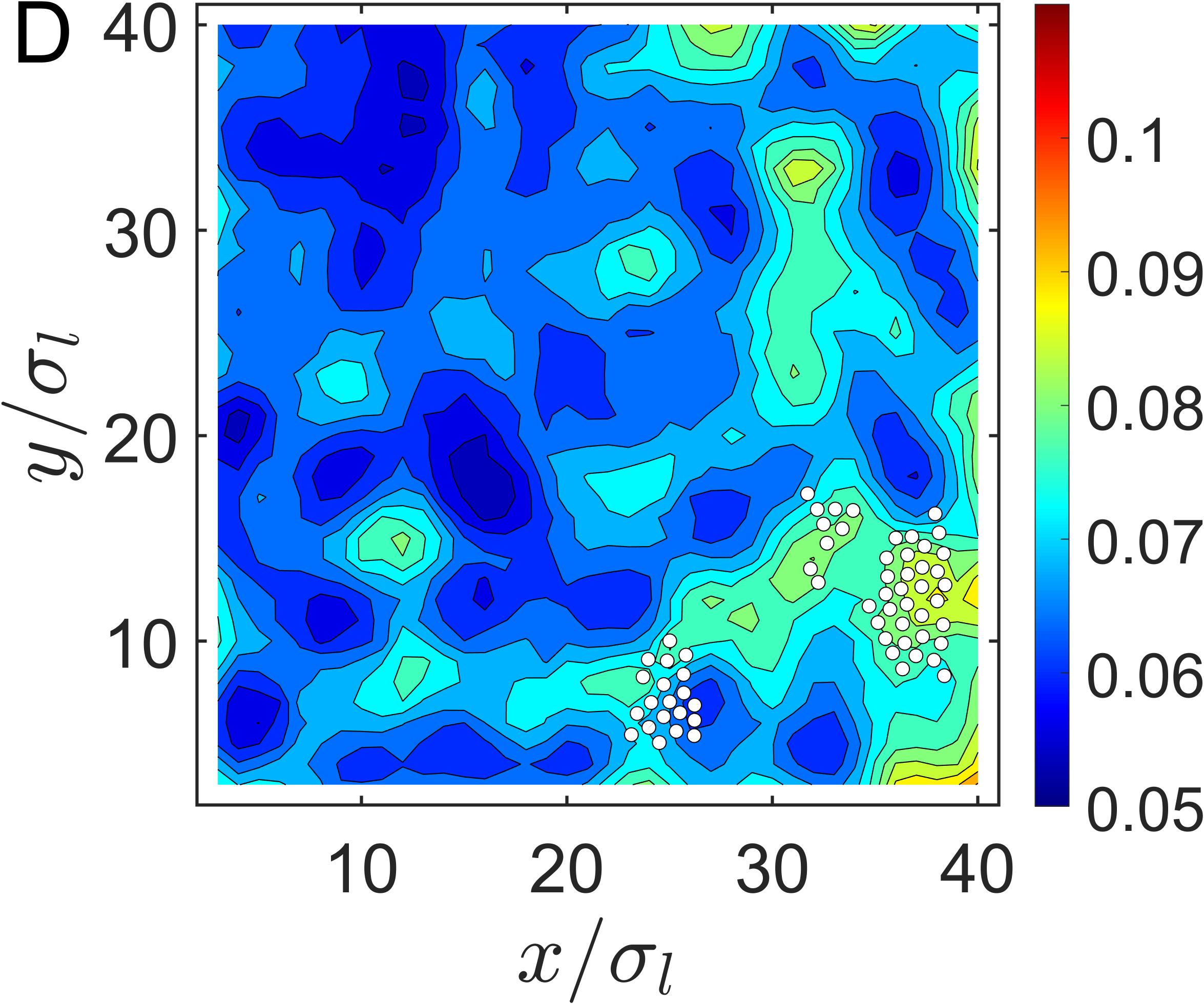}&
\includegraphics[width=.29\textwidth]{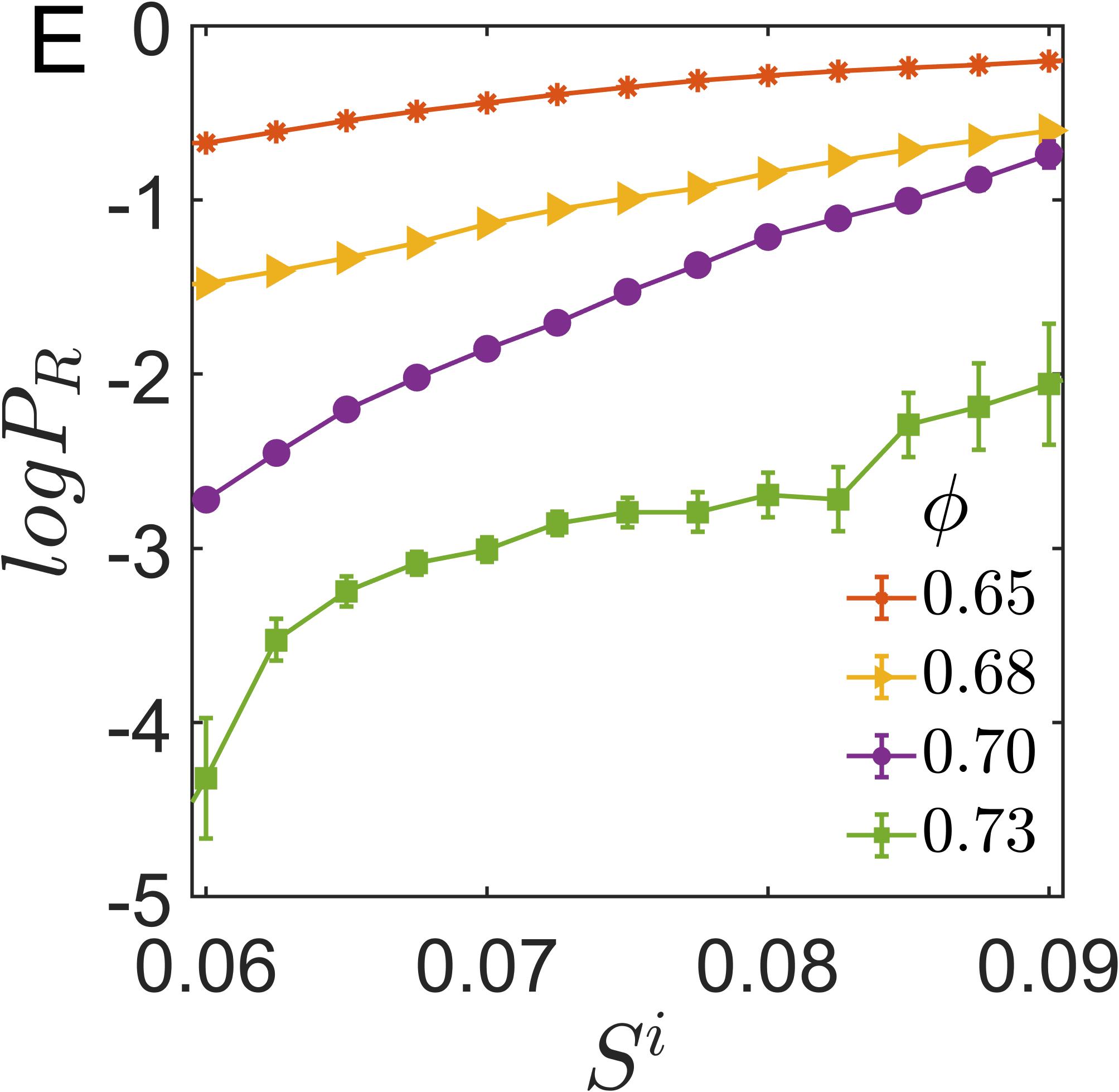}&
\includegraphics[width=.295\textwidth]{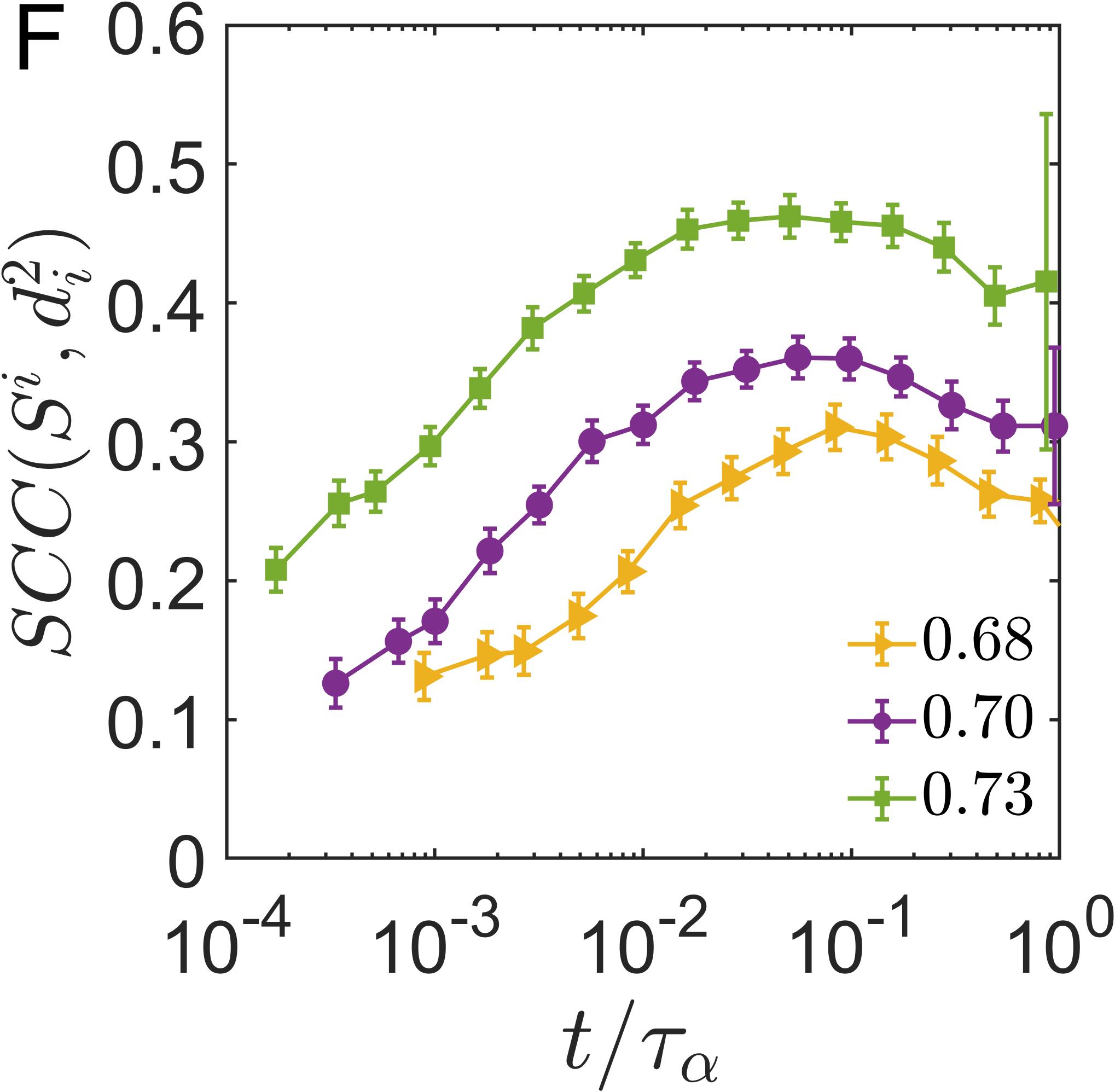}\\
\end{tabular}
\caption{Structure-dynamics correlations in 2D bidisperse suspensions. (A) A reconstruction of the local structural order parameter of the system at $\phi=0.70$. The particles are color coded based on the magnitude of $S^i$. (B) The distributions of structural order parameter $P(S^i)$ at various area fractions of the colloids. Different symbols are used to distinguish the area fractions. Bottom inset: The average SOP $\bar{S^i}$ of the distributions as a function of $\phi$. Top inset: The particle level radial distribution functions $g(r)$ averaged over hard and soft particles are determined separately. Particles with $S^i$ below the mean $\bar{S^i}$ are considered hard and vice-versa for soft particles. (C) The distribution $P(S^i)$ of rearranging particles (square) and all particles (triangle). The area under the shaded region gives the fraction of rearranging particles with SOP greater than the peak of the distribution $P(S^i)$. (D) The rearranging particles (white circles) are overlapped on the structural order parameter contour of a reference configuration at $\phi=0.70$. (E) The probability that a particle with structural order parameter $S^i$ undergoes rearrangement $P_R$ at various area-fractions ranging from $\phi=0.65-0.73$. (F) The Spearman rank correlation coefficient (SCC), which correlates the rank of $S^i$ and the rank of relative displacement of a particle $d_i^2(t/\tau_{\alpha})$, is shown as the function of time (Eq.4 in Materials and Methods) (C)-(E) The rearrangements are identified by tracking the particles' motion over a time scale $\tau_r=100s$ and those having $d_i^2>l_r^2=0.25$ (see section F in Materials and Methods). (E)-(F) the error bars are equal to the standard deviation. All the statistical analysis is done using large particles.}
\label{Fig2}
\end{figure*}
%Note that the results in panels (B)-(D) and (F) were obtained by considering only the large particles.

\section*{Structural relaxation due to thermal fluctuations}

We next apply this order parameter to a colloidal glass. We prepare an amorphous monolayer of bidisperse colloidal suspensions over a range of densities by mixing silica particles with diameter $2.32\mu m$ and $3.34 \mu m$ in a $1:1$ ratio. The expression of the SOP for the binary system is given in the materials and methods section. A reconstruction of the colloidal particles at $\phi=0.70$, with particles color-coded according to their structural order parameter, is depicted in Fig. \ref{Fig2}\emph{A}. In this representation, the SOP of particles has been coarse-grained, see section I-C in the supporting information. The image shows interpenetrating regions of high and low SOP, offering a direct visualization of the structural heterogeneity within the amorphous system. To gain a quantitative understanding, we plot the distributions of the order parameter for different particle densities in Fig. \ref{Fig2}\emph{B}. The distribution shifts to the left with increasing density, indicating that the local particle environments become harder. This is also apparent from the bottom inset, where the average SOP, $\bar{S^i}$, is observed to decrease with $\phi$, implying an increasing depth of the caging potential. Concomitantly, there is a simultaneous increase in the relaxation time of the system (see Fig. S1C in supporting information); therefore, the structure factor, on an ensemble level, captures the proliferation of dynamic relaxation of the system. We can now utilize the order parameter to identify structural hallmarks of hard and soft environments. Hard environments encompass particles with $S^i$ less than the mean of the distribution, while soft environments encompass particles with $S^i$ greater than the mean. A comparison of the pair correlations (RDF) for these two sub-populations is depicted in the upper inset of Fig. \ref{Fig2}\emph{B}. Notably, the first peak of $g(r)$ is more pronounced for hard particles, and a similar effect is observed for other higher-order peaks. Therefore, hard particles have a more structured neighborhood than soft particles. The SOP effectively identifies these distinct structural environments. \\

To link the structure and dynamics on a local level, we compare the SOP distribution of rearranging particles with that of all particles, where rearranging particles are identified from their relative displacement $d_i(\Delta t)$ over a time scale $\tau_r$ (see section F in Materials and Methods for details). Figure \ref{Fig2}\emph{C} shows that the distribution for rearranging particles, $P(S^i|R)$, is shifted to the right, indicating they are associated with a softer neighborhood. The shaded area in Fig.\ref{Fig2}\emph{C} represents the fraction of rearranging particles whose SOP is larger than the average SOP of all particles, which is nearly eighty-seven percent. This indicates that rearranging particles is indeed associated with a larger SOP.\\

To establish a more direct correlation between the SOP and particle dynamics, we plot the fraction of particles $P_R(S^i)$ undergoing rearrangement as a function of $S^i$ in Fig.\ref{Fig2}\emph{E}, see section H in Materials and Methods for details. The panel shows SOP values in the range $S=0.06~-~0.09$, where the distribution $P(S)$ is finite at all $\phi=0.65 - 0.73$. Clearly, the probability of rearrangement grows with $S^i$, and this growth is more pronounced at higher densities. The form of the $P_R(S)$ is dependent on the values of $\tau_r$ and $l_r$ that are chosen to detect the rearrangements. A detailed discussion on the effect of thresholding on $P_R(S)$ is presented in section III(C) of SI. These results confirm that dynamics is correlated to the structural order parameter, especially at higher densities where there is a better decoupling of the short and long time dynamics, and the cage around a particle becomes longer lived. So even if the SOP values at two densities are similar, at higher density, the longer lived cage will have a stronger effect on the dynamics. This is in a way similar to stronger correlations between structure and dynamics at lower temperatures found in numerical simulations of molecular systems \cite{Liu15, Bhattacharyya22}. \\

A visual impression of these correlations is presented in Fig.\ref{Fig2}\emph{D}, where the contours are drawn using $S^i$ of particles in a reference configuration, and the rearranging particles are shown as white circles. The rearranging particles are indeed located on top of regions of high SOP, visually establishing a close correlation between the dynamics and the structure characterized by the SOP. These observations are further confirmed by direct correlation analysis, by computing the Spearman rank correlation coefficient (SCC) between the particles' SOP value, $S^{i}$, and their relative displacements as a function of scaled time, see section I in Materials and Methods for details. For all area fractions in Fig.~2\emph{F}, the correlation initially grows, and it decays slowly with increasing time. \\

The maximum Spearman rank correlation coefficient (SCC) found in our experiments is $\sim 0.45$, observed at $\phi=0.73$. This value is higher than those reported in earlier experimental studies on structure-dynamics correlations in colloidal systems, which found values around 0.3 \cite{Chen16, Chen19}. However, it is lower than the correlation coefficients reported in recent machine learning studies using simulated data, where the Pearson correlation coefficient between structure and propensity ranged from $0.6-0.7$ under quiescent conditions \cite{Kohli20}. These values were obtained using graph neural network methods. Other methods, including the support vector method was found to yield smaller values of Pearson correlation coefficient. The magnitude of the SCC is also influenced by the coarse-graining length (L) \cite{Tanaka18-1, Tanaka19, Tanaka20, Bhattacharyya23, Coslovich20} used in the calculation of the structural order parameter (see Eq. 5 and Eq. 6 in section J of Materials and Methods). There is an optimal value of L that further improves these correlations. A detailed discussion is presented in section IV(A) of the supplementary information (SI). It is important to note that simulations under quiescent conditions typically use iso-configurational runs to average out the effect of initial particle velocities. Implementing these protocols in experiments is not feasible, so we expect that the correlation coefficients in experiments will be lower than those in simulations.

%\textcolor{red}{The maximum Spearman rank correlation coefficient (SCC) found in our experiments is $\sim 0.45$, observed at $\phi=0.73$. This value is higher than those reported in earlier experimental studies on structure-dynamics correlations in colloidal systems, which found values around 0.3 \cite{Chen16, Chen19}. However, it is lower than the correlation coefficients reported in recent machine learning studies using simulated data, where the Pearson correlation coefficient between structure and propensity ranged from $0.5-0.65$ under quiescent conditions \cite{Kohli20}. The magnitude of the SCC is also influenced by the coarse-graining length (L) \cite{Tanaka18-1, Tanaka19, Tanaka20, Bhattacharyya23, Coslovich20} used in the calculation of the structural order parameter (see Eq. 5 and Eq. 6 in Section J of Materials and Methods). There is an optimal value of L that further improves these correlations. A detailed discussion is presented in Section IV(A) of the supplementary information (SI). It is important to note that simulations under quiescent conditions typically use iso-configurational runs to average out the effect of initial particle velocities. Implementing these protocols in experiments is not feasible, so we expect that the correlation coefficients in experiments will be lower than those in simulations.}

\section*{Structural relaxation under shear}

\begin{figure*}[t!]
\centering
\begin{tabular}{lll}
\includegraphics[width=.31\textwidth]{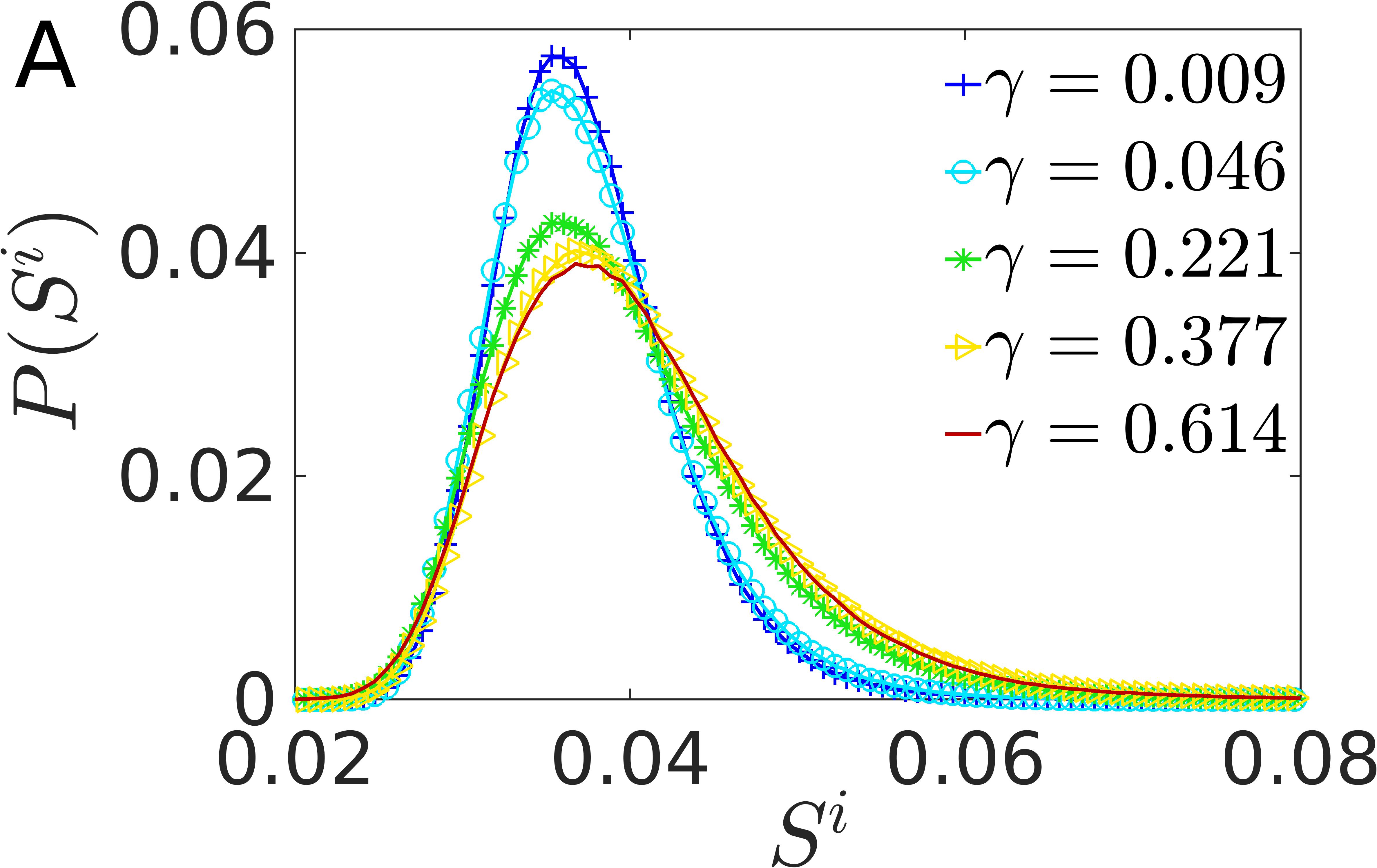}&
\includegraphics[width=.33\textwidth]{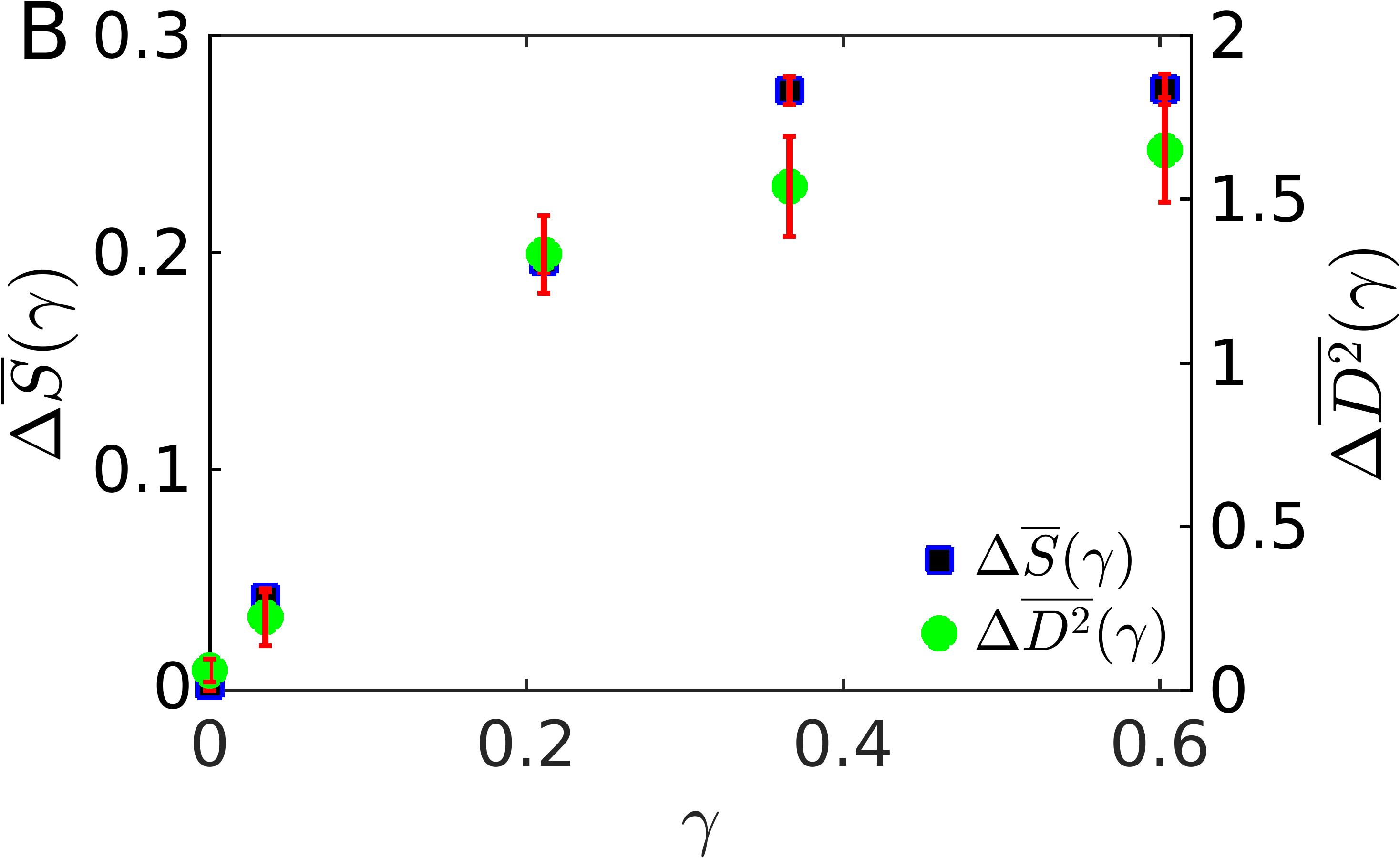}&
\includegraphics[width=.30\textwidth]{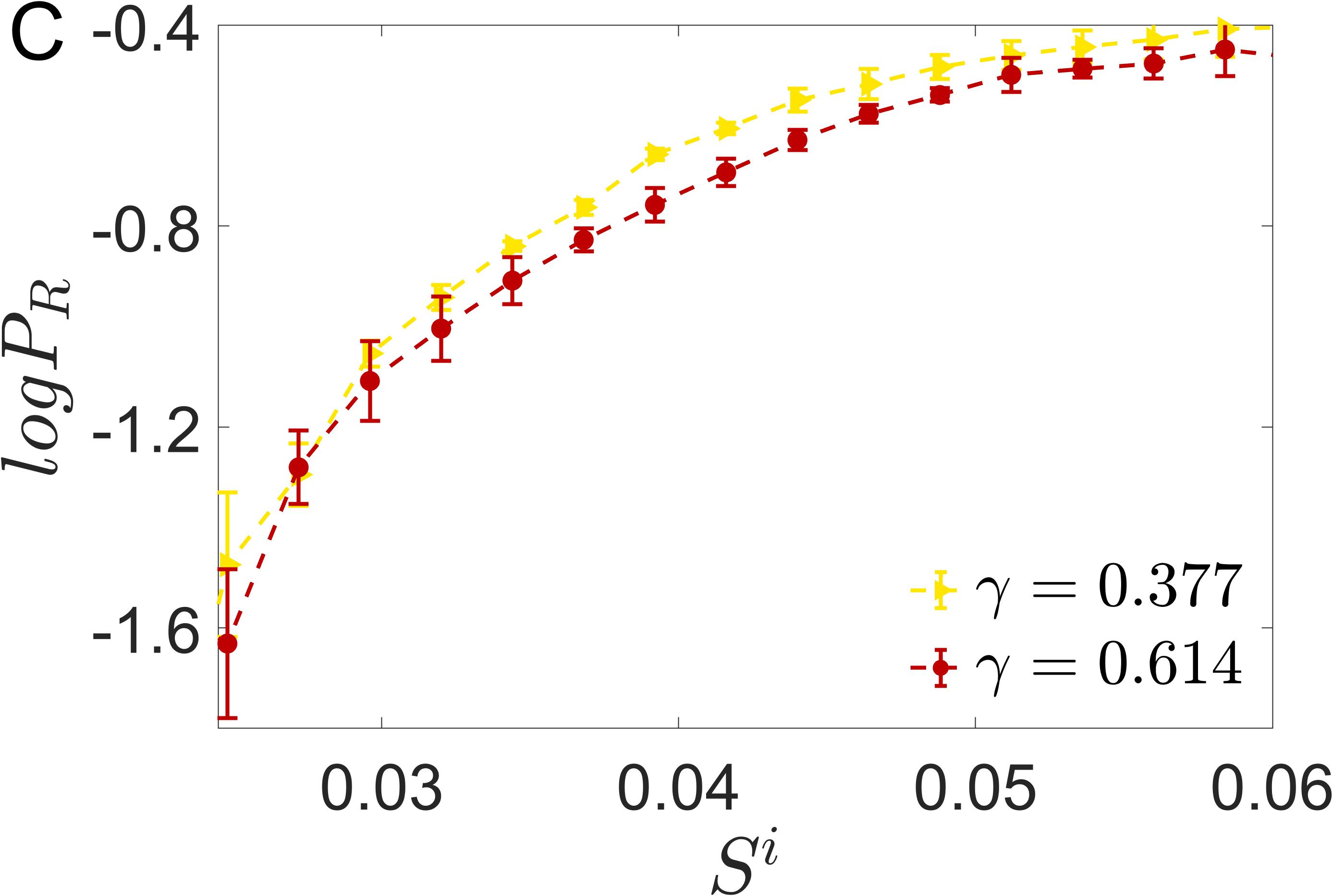}\\
\includegraphics[width=.30\textwidth]{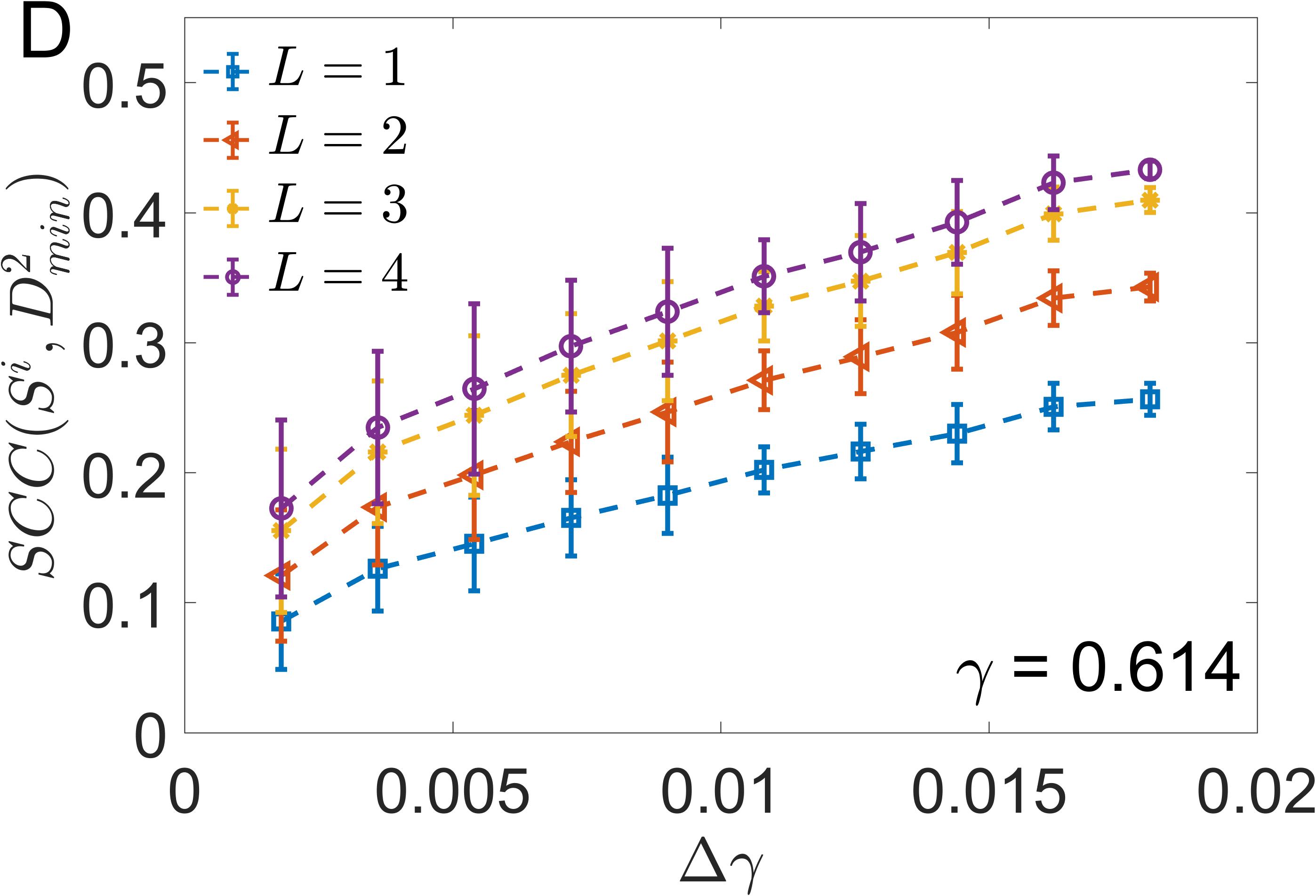}&
\includegraphics[width=.31\textwidth]{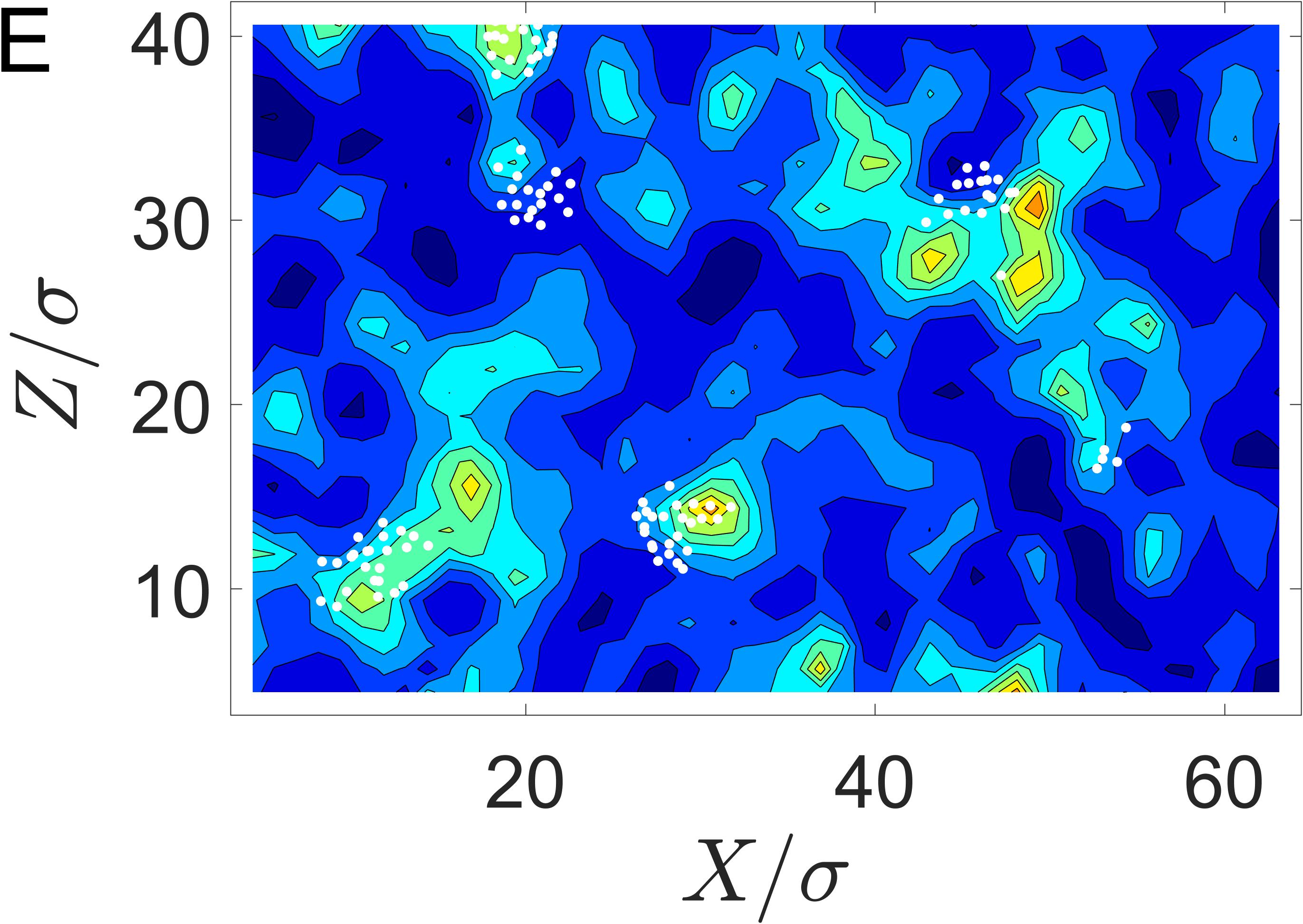}&
\includegraphics[width=.34\textwidth]{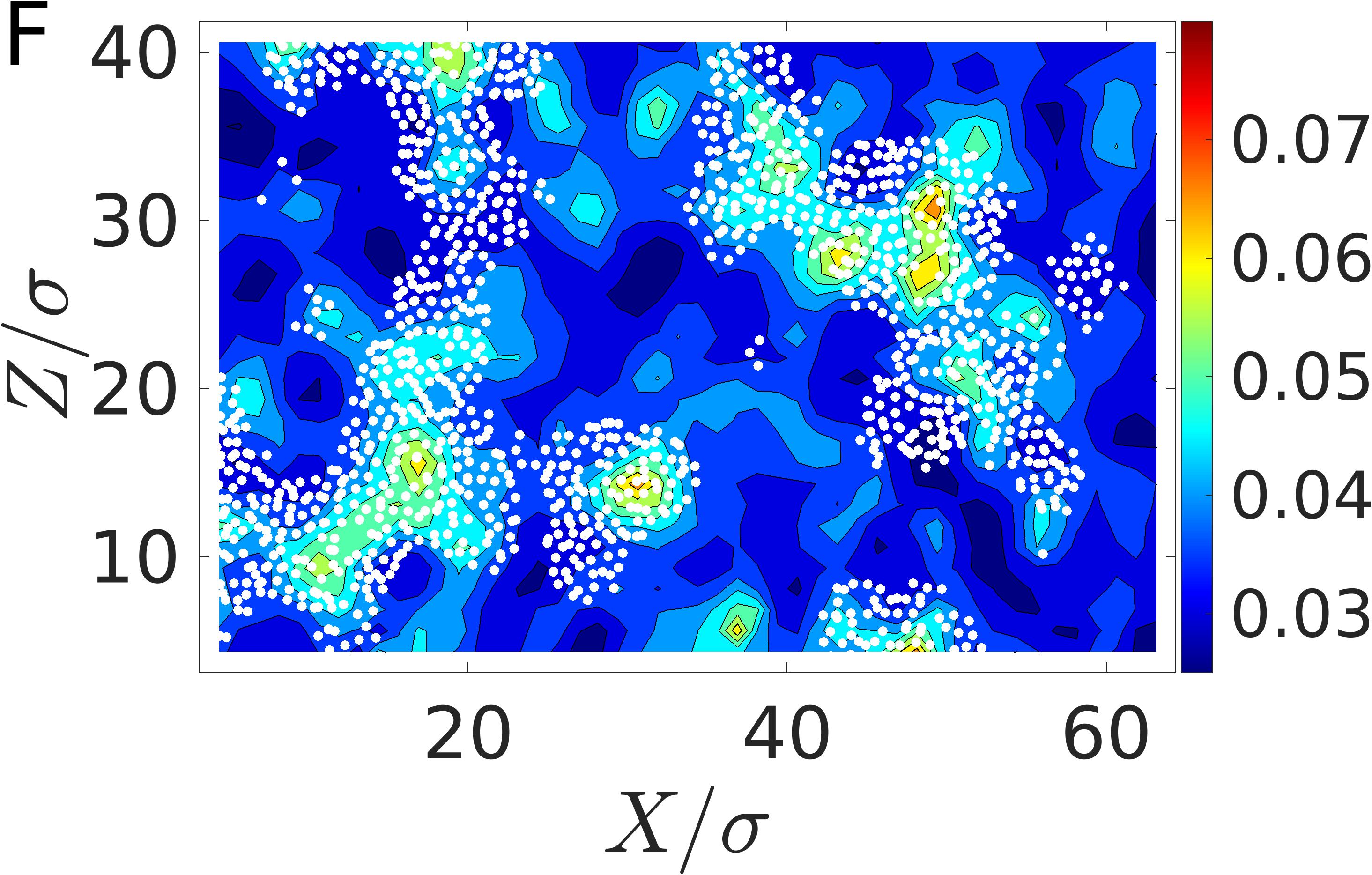}\\
\end{tabular}
\caption{Structural order parameter and plastic events in dense colloidal suspensions under shear at a constant $\dot{\gamma}= 1.5\times 10^{-5}s^{-1}$. 
(A) The effect of applied shear on the distribution of local order parameter, $P(S^i)$, for strain values ranging from $\gamma\sim0-0.6$. The distributions are obtained by averaging over multiple configurations centred around $\gamma$. 
(B) Normalised change in the average SOP, $ \Delta \bar{S}(\gamma) = \left< \frac{\bar{S}(\gamma)-\bar{S}(\gamma_{0})}{\bar{S}(\gamma_{0})} \right>$, is presented on the left-side vertical axis using square symbols, where the bar represents averaging over all particles and the angular bracket denotes averaging over multiple configurations centred around $\gamma$. The constant $\gamma_0$ is the smallest strain value considered in our study. The scaled plastic deformation $\Delta \bar{D^2}(\gamma) = \left< \frac{ \bar{D^2}_{min} (\gamma,\gamma_0)-\bar{D^2}_{min} (0,\gamma_{0})} {\bar{D^2}_{min}(0,\gamma_{0})} \right>$ is presented on the right side y-axis using circles. 
(C) The probability of rearrangement of a particle with order parameter $S^i$ for two different strain values $\gamma =0.377$ and $0.614$ in the steady state. The $P_R$ is computed over a strain interval $\Delta \gamma_r$, and it is averaged over several such instances in the steady state. (D)  The Spearman rank correlation between the SOP and the $D^2_{min}$ in the steady state at $\gamma=0.614$. Different symbols represent the varying coarse-graining length scales of SOP. The error bars in panels (B)-(D) are equal to the standard deviation. (E)-(F)The SOP of a reference configuration in the steady state at $\gamma=0.614$ is shown as contours, and the particles that undergo plastic rearrangements in a small section of two particle diameters thick are shown in white circles. The rearrangements are identified over strain windows $\Delta \gamma=0.0045$ (E) and $\Delta \gamma=0.018$ (F) using $D_r$ as the threshold on non-affine displacements of particles.} 
\label{Fig3}
\end{figure*}

We find that the structural order parameter is also a reliable structural indicator for predicting shear transformation under applied shear. We use dense colloidal suspensions in 3D under plane shear at a constant shear rate of $\dot{\gamma} \sim 1.5 \times 10^{-5} s^{-1}$; this shear rate is of the order of the inverse relaxation time of the system, indicating the system is weakly driven \cite{Schall17}. The particles have a diameter of $1.4 \mu m$, with a polydispersity of $7\%$, see materials and methods for details. \\

Upon start up of the shear, the order parameter distribution shifts to the right, as shown in Figure \ref{Fig3}\emph{A}. Especially, the right wing of the distribution shifts to higher values, indicating that soft regions become even softer upon the application of shear. In this representation, we have averaged distributions over multiple configurations centred around $\gamma$, and we have treated the system isotropically in the calculation of the SOP, although the system is strained. We quantify this trend by plotting changes $\Delta \bar{S}(\gamma) = \left< \frac{\bar{S}(\gamma)-\bar{S}(\gamma_{0})}{\bar{S}(\gamma_{0})} \right>$ of the average SOP, $\bar{S}(\gamma)$, with respect to its initial value in Figure \ref{Fig3}\emph{B}. Here, $\gamma_0=0.0045$ represents the smallest strain interval considered in our study, and the bar and angular brackets represent averaging over all particles and multiple configurations, respectively. Initially, $\Delta \bar{S}$ increases with applied strain, suggesting a reduction in the average depth of caging potential experienced by the particles and consequent shear softening of the system. Beyond a certain strain threshold, the system attains a steady state with no overall structural changes. This trend is similar to the one we observed for the non-affine displacements \cite{Schall17}, indicating a coupling of the structure and plasticity. To show this, we overlay the scaled change of the nonaffine displacements, $\Delta \bar{D^2}(\gamma)$ in Figure \ref{Fig3}\emph{B} (green dots and right-hand vertical axis). The nonaffine displacements and structural order parameter show a remarkably similar trend. Both increase with strain during the transient stages of deformation and saturate after the system yields and attains a steady state. This trend persists at higher shear rates; see Fig.S15 in SI for a similar analysis at $\dot{\gamma}=10^{-4} s^{-1}$.\\

Earlier investigations into sheared colloidal suspensions based on mode coupling theory were successful in relating the shear stress to microscopic structure \cite{Cates09, Fuchs13, Schall13, Laurati18, Petekidis12, Egelhaaf12}. The rheological stress was also reported to be related to the elastic free energy consisting of affine and non-affine contributions \cite{Zaccone17}. In addition, oscillatory shear measurements of 2D amorphous solids \cite{Arratia20} investigated the correlation between relaxation rates of plastic flow and the structure quantified via the excess entropy. However, these studies did not explore the correlation between local structure and plastic events under shear. \\

To establish strong correlations between the structure and plastic events, we look at the rearrangement probability of a particle as a function of its SOP value, $P_R(S^i)$, in a way similar to the probability of relaxation we used before for the quiescent system. In the sheared case, rearranging particles are identified as those having non-affine displacements larger than a threshold $D^2_r$ in a strain interval $\Delta\gamma_r$ (see section G in Material and Methods for details). During the initial stages, at small $\gamma$, the deformation is largely elastic, with limited plastic deformation \cite{Procaccia10}, as evidenced by merely a small number of rearrangements within a typical observation window of $\Delta\gamma_r=0.016$. This is in line with the overall small nonaffine displacement at small strains as shown in Fig.~\ref{Fig3}\emph{B} and Fig.5 in Materials and Methods. We, therefore, focus on higher strains approaching the steady state, for which we show the probability of rearrangement as a function of the structural order parameter in Fig. \ref{Fig3}\emph{C}. Consistent with experiments under quiescent conditions, we observe a positive correlation between rearrangements and SOP, thus affirming that particles with larger SOP are more likely to rearrange. \\

For further evidence, we present the Spearman rank correlation analysis in Fig.~3\emph{D}, see section I in Materials and Methods for details. This analysis shows the correlation coefficient as a function of strain increment $\Delta \gamma$ in the steady state, corresponding to a total strain $\gamma=0.614$. The results clearly indicate that correlations grow with $\Delta \gamma$. However, due to the finite duration of image acquisition, the large strain behavior is not captured in the experiments. It is to be noted that the total duration of the experiment exceeds 12 hours at $\dot{\gamma}= 1.5\times 10^{-5}s^{-1}$, so the images are acquired in small strain intervals. The different curves in the plot illustrate the effect of coarse-graining length (L). As L increases, so do the correlations, with the maximum correlation coefficient from our experiments reaching approximately 0.45. Few simulations have investigated these aspects. Recent machine learning studies \cite{Kohli20} of sheared systems using an athermal quasi-static protocol measured structure-dynamics correlations by calculating the Pearson correlation coefficient, yielding values in the range of $0.5-0.6$ using graph neural network method. The support vector methods was found to give smaller values of correlation coefficient in the range $0.25-0.35$. These simulations did not explore the effects of finite temperature and finite shear rates, which are closer to experimental conditions. In light of these studies, the correlation coefficients measured in our experiments are comparable to simulations.

These correlations between the structural order parameter and plastic rearrangements are clearly visible in panels \ref{Fig3}\emph{E} to \ref{Fig3}\emph{F}, showing deformation in the steady state. The panels depict the particles undergoing plastic rearrangements with increasing strain intervals as white dots overlaid on a contour plot of the SOP of the initial configuration. As the system is sheared, plastic events cluster in regions of large SOP. These maps are direct visual evidence of the correlation between structure and dynamics. An earlier study \cite{Liu17} using ML techniques elucidated the universality of yield strain in a large class of amorphous solids by establishing a correlation to structural softness. Our study suggests that these correlations extend beyond yielding into the steady-state flow and establish a rational structural order parameter directly derived from the short-range order of the amorphous material. \\

\section*{Concluding remarks}
Our findings demonstrate that the relaxation of dense amorphous colloidal suspensions, whether due to thermal fluctuations or weak applied shear, has a structural origin. We have elucidated this phenomenon using a local structural order parameter (SOP), which is the inverse of the local caging potential experienced by each particle due to its neighboring particles. The particles with large SOP are associated with loosely packed neighborhoods and the investigation of rearrangement probability $P_R(S)$ and the Spearman rank correlation coefficient confirm strong structure-dynamics correlations. The magnitudes of the SCC measured in our experiments compare well with earlier studies. These evidences point to the effectiveness of SOP in identifying localized defect-like regions in amorphous suspensions susceptible to particle rearrangements, which leads to relaxation. Unlike in crystalline materials, where the dislocations or grain boundaries are extended objects, the weak-defective regions in amorphous materials appear to be localized. Thus, our study provides a fresh perspective on the plasticity of disordered solids centred on structural heterogeneity characterized by the effective caging potential. These insights could further our understanding of the structural role in the shear banding instabilities of complex fluids.
%These insights could further our understanding of the structural role of relaxations in glasses and shear deformations and shear banding instabilities of complex fluids.

\section*{Materials and Methods}
\subsection{Experimental realization of 2D colloidal crystals} 
The monolayer of a 2D colloidal crystal was made by sedimenting silica beads of diameter $\sigma=3.34 \mu m$ in deionized(DI) water. The gravitational height of the particles was measured to be $0.02 \mu m$, which is smaller than the particle size. The 2D crystals thus formed contain vacancies and grain boundaries. The particles are imaged using bright-field microscopy with a field of view of $280*280 \mu m$. The features were found using particle tracking algorithm\cite{Grier96} and $4866$ particles are obtained within the field of view.\\

\subsection{Experimental realization of 3D colloidal crystals} 
The 3D crystal was created by suspending silica beads of $\sigma= 1 \mu m$ in a $80:20$ mixture of glycerol and DI water to match the refractive index. To visualise the particles in 3D, a $1$ mM concentration of Rhodamine-6G dye was added to make the solvent fluorescent. The sample was left on the microscope for $\sim 24h$ prior to measurement so that crystals formed with visible grains. The imaging was done using Leica-Dmi8 confocal microscope with a field of view of $62*62*32 \mu m$ containing nearly $1,50,000$ particles. \\

\subsection{Experimental realization of 2D amorphous suspensions} 
We have used a 50:50 binary mixture of silica colloids of diameters $\sigma_l = 3.34 \mu m$ and $\sigma_s = 2.32 \mu m$ and a size ratio $\sigma_l/\sigma_s=1.4$. The colloids are dispersed in DI water and loaded into a thin chamber with a thickness of $100 \mu m$, which was created by sandwiching two coverslips using double-sided tape. The coverslips were plasma-cleaned to prevent the colloids from sticking to the glass surface. The particles were allowed to sediment under gravity to form an amorphous monolayer with a gravitational height of $0.05 \mu m$ and $0.02 \mu m$ for the small and large particles, respectively. The sample was left on the microscope for approximately 2 hours before imaging. The number density of the colloids was varied to adjust their area fraction in the monolayer, and they were imaged in a region measuring $145\mu m \times 145 \mu m$ at a frame rate of 21 frames per second. The trajectories of individual particles were determined using a standard particle tracking algorithm \cite{Grier96}. In this study, we investigated a range of area fractions from $\phi=0.65 - 0.73$, which contains approximately $1800$ to $2300$ particles in the field of view.\\

\subsection{Experimental techniques used for performing shear measurements} 
The shear experiments are performed using dense colloidal suspensions of sterically stabilized fluorescent polymethylmethacrylate particles in a density and refractive index matching mixture of cycloheptyl bromide and cis-decalin. The particles have a diameter, $\sigma = 1.3 \mu m$ and a polydisperity of $7\%$ to prevent crystallization. The suspension was centrifuged at an elevated temperature to obtain a dense sediment, which was subsequently diluted to get a suspension of the desired volume fraction $\phi\sim0.60$. The sample was sheared using a shear cell with two parallel boundaries separated by a distance of $\sim 50\sigma$ along the $z-$direction \cite{Schall11}. A piezoelectric device was used to move the top boundary in the $ x-$ direction to apply a shear rate of $1.5 \times 10^{-5}$. To prevent boundary-induced crystallization in our samples, the boundaries were coated with a layer of polydisperse particles. Confocal microscopy was used to image the individual particles and to determine their positions in three dimensions with an accuracy of $0.03 \mu m$ in the horizontal and $0.05 \mu m$ in the vertical direction. We tracked the motion of $\sim 2 \times 10^{5}$ particles during a $25$-min time interval by acquiring image stacks every $60~s$. The data was acquired during a small observation window at various strain values $\gamma$.
%All the measurements presented here were made in the steady state, after the sample had been strained to $>50\%$ at shear rate of $1.5\times 10^{-5}s^{-1}$. \\

\subsection{Structural order parameter of bidisperse suspensions} 
The local structural order parameter is the inverse depth of the local caging potential, which is expressed in real space as $S^i$ = $\frac{1}{\beta \Phi^{i}(\Delta r=0)}$. The local caging potential of a bidisperse system is written as
\begin{equation}\label{eq:2}
  \beta \Phi^{i}_{1}(\Delta r=0) = -\rho_v \int \mathbf{dr}\sum_{v=1}^2~x_{v}g^i_{1v}(r)[g^i_{1v}(r)-1],
\end{equation}
where $g^i_{1v}(r)$ is the mollified particle level pair function for type $1$ particle\cite{Parrinello17_2}. It is expressed as 
\begin{equation}\label{eq:3}
g_{1v}^{i}(r)=\frac{1}{ \rho_v \mathbf{dr}}\sum_j\frac{1}{\sqrt{2\pi \delta^2}}\text{e}^-\frac{(r-r_{ij})^2}{2\delta^2},
\end{equation}
where $\delta$ is the Gaussian broadening factor that makes local $g_{uv}^{i}(r)$ continuous. The values of the broadening parameter in 2D and 3D systems were $\delta =0.06$ and $0.02$, and $\mathbf{dr}=2\pi r dr \hspace{2mm}\text{and} \hspace{2mm}4\pi r^2 dr$ respectively. For details see section IC of SI.\\

\begin{figure}[h!]
  \centering
    \includegraphics[width=0.35\linewidth]{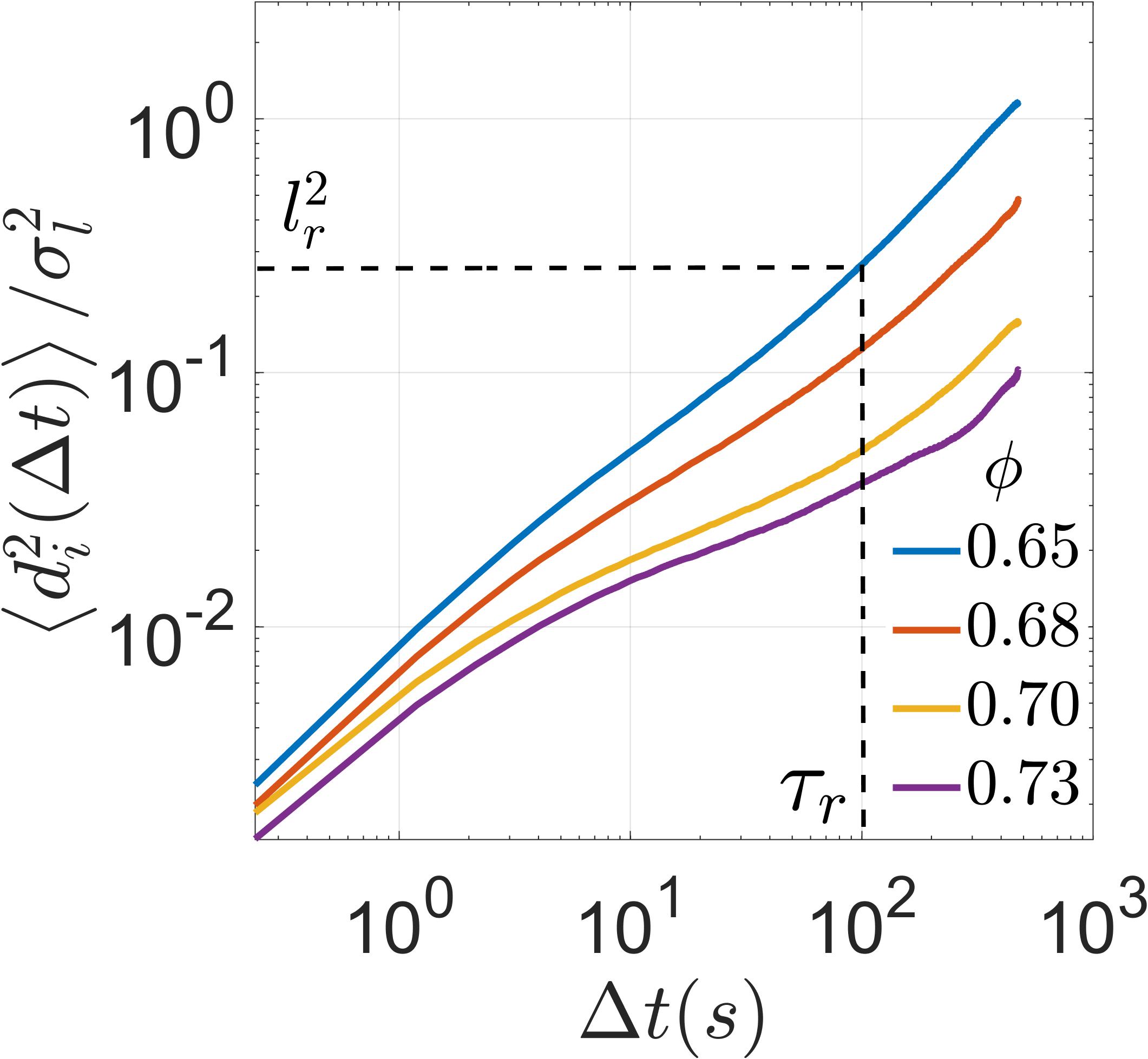}
  \caption{Mean square relative displacement over a range of area-fractions from $\phi=0.65-0.73$. The onset of diffusive motion at $\phi=0.65$ is marked using the black dashed vertical line, which corresponds to $\tau_r=100s$. The horizontal dashed line represents the threshold displacement $l_r=0.5$. These values of $\tau_r$ and $l_r$ are fixed and are used for identifying rearrangements at all area-fractions.}
\label{Fig4}
\end{figure}

\subsection{Identifying rearrangements in quiescent suspensions} 
We define rearrangements in our 2D measurements based on the relative displacements \cite{Bonn11, Weeks17} of particles, which is given by the following expression: 
\begin{equation}
d_i(\Delta t)=\left[\frac{1}{n} \sum_{j=1}^n\left|\Delta \mathbf{r}_{i j}(t+\Delta t)-\Delta \mathbf{r}_{i j}(t)\right|^2\right]^{1/2},
\end{equation}
where $\Delta \mathbf{r}_{ij}$ is the vector joining the centers of particles $i$ and $j$, and $n$ is the number of nearest neighbors within a cutoff distance equal to the first minima of $g(r)$. The mean square relative displacement is shown in Fig.~\ref{Fig4} for a range of area-fractions from $\phi=0.65-0.73$. The black vertical line marks the timescale for the particles to transition from subdiffusive to diffusive motion at $\phi=0.65$. This corresponds to a timescale of $\tau_r=100s$ and a normalized length scale of $l_r=0.5$. We use these values to determine rearrangements at all area-fractions in quiescent measurements. Any particle that has normalized relative displacement more than $l_r$ on the time scale $\tau_r$ is said to have rearranged. \\

\subsection{Identifying rearrangements in suspensions under shear} 
\begin{figure}[h!]
  \centering
    \includegraphics[width=0.35\linewidth]{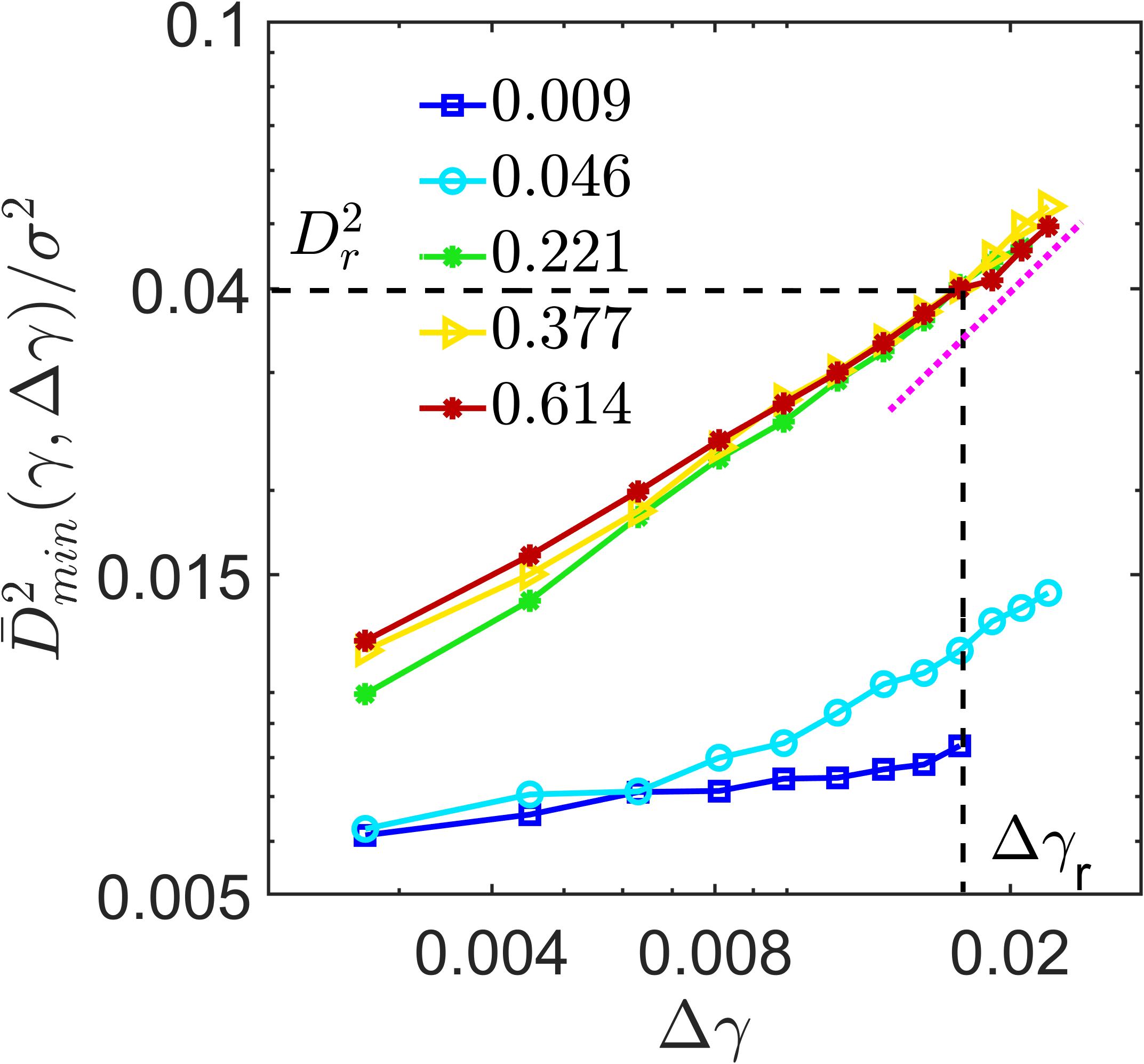}
  \caption{The average non-affine displacements of particles $\bar{D}^2_{min}(\gamma,\Delta \gamma)$ for the sheared system at various stages of macroscopic deformation. The different curves correspond to different values of macroscopic strain $\gamma$, which is denoted as legend labels. The strain increment $\Delta \gamma$ is measured with respect to the total macroscopic strain $\gamma$. The $\bar{D}^2_{min}(\gamma,\Delta \gamma)$ values are scaled by the particle size. The black vertical dotted line represents the smallest strain window $\gamma_0$. The black vertical and horizontal dashed line indicates the strain interval $\Delta\gamma_r$ and threshold non-affine displacement $D^2_r$, respectively, used for identifying rearrangements in the sheared system. The  magnitudes of $\Delta\gamma_r=0.016$ and $D^2_r=0.04$. All particles with $D^2_{min,i}(\gamma,\Delta\gamma_r)>D^2_r$ are identified as rearrangements over a strain scale $\Delta\gamma_r$. The magenta dotted line with a slope unity marks the onset of diffusive motion. }
\label{Fig5}
\end{figure}

We identify regions of plastic deformation in sheared suspensions by examining the nonaffine displacements of particles \cite{Langer98, Schall11}. For a strain increment $\Delta \gamma$, measured with respect of total strain $\gamma$, the non-affine displacement of a tagged particle is defined as $D^2_{min,i}(\gamma,\Delta\gamma)=  \frac{1}{N_i}\sum_{j=1}^{N_i} \left[\mathbf{r}^{ij}(\gamma+\Delta \gamma)- \mathbf{\Gamma}_j(\gamma) \mathbf{r}^{ij}(\gamma)\right]^2$. Here, \emph{i} is the index of tagged particle, $\mathbf{r}^{ij}$ is the displacement vector between particle $i$ and its nearest neighbors $j$, $N_i$ is the number of first nearest neighbors of particle $i$ based on the first minima of $g(r)$, and $\mathbf{\Gamma}_j$ is the best-fit affine deformation tensor that minimizes $D^2_{min,i}$. \\

A plot of average non-affine displacements $\bar{D}^2_{min}(\gamma,\Delta \gamma)=\frac{1}{N}\sum_{i=1}^{N}D^2_{min,i}(\gamma,\Delta\gamma)$ is shown in Fig.\ref{Fig5} at various stages of macroscopic deformation, represented by $\gamma$. The magenta dotted line shows a line of slope $1$. The black dotted vertical line indicates $\gamma_0$, the smallest strain interval considered in our study. The black vertical dashed line corresponds to a strain interval $\Delta\gamma_r$ that marks the transition from sub-diffusive to diffusive motion. The non-affine displacement $D^2_r$ is the threshold value for identifying rearrangements. The rearranging particles are those that have $D^2_{min,i}(\gamma,\Delta\gamma_r)>D^2_r$ on a strain scale $\Delta\gamma_r$. In our calculation $\Delta\gamma_r=0.016$ and $D^2_r=0.04$. Note that the average non-affine displacements of particles, $\bar{D}^2_{min}(\gamma,\Delta \gamma)$, in the transient or early stages of deformation is small, which implies that the deformation in predominantly elastic with a small number of rearrangements. However, once the system attains a steady state, the average non-affine displacement and the number of plastic rearrangements increase dramatically.

\subsection{Calculating the rearrangement probability $P_R$}
To calculate the rearrangement probability, denoted as $P_R$, we first determine the number of particles with SOP in a small interval around $S^i$. Next, the number of particles undergoing rearrangements is determined. The fraction of particles that have rearranged as a function of $S^i$ is the rearrangement probability $P_R$.\\

The method used to average and obtain the $P_R(S)$ curves is described below. For quiescent measurements, we captured 10,000 images of 2D binary systems at 21 frames per second over a total duration of 476 seconds. To identify rearrangements, we utilized a time window of $\tau_r = 100$ seconds and $l_r^2 = 0.25$. The SOP values of all particles in all images formed a large dataset. To calculate $P_R(S)$ and estimate the error bars, we created 100 datasets from this pool using the Fisher-Yates shuffle algorithm. Each dataset consisted of approximately 5\% of the values. The final $P_R(S)$ was obtained by averaging these 100 datasets, with the error bars representing the standard deviation. The bin width of $S^i$ in both quiescent and sheared systems was set at $0.0025$. The effectiveness of this method was compared to a second method of calculating $P_R(S)$, with details provided in section III(A) of the supplementary information (SI). These methods ensure sufficient statistics for calculating $P_R(S)$, which is crucial at higher area fractions where the number of rearrangements is small.\\

For sheared systems, where the number of rearranging particles is large, $P_R(S)$ is determined separately for each configuration and then averaged. As the shear measurements are conducted with bulk 3D colloidal samples, the number of particles is significantly larger compared to the 2D binary system used in quiescent measurements.\\

\subsection{Calculation of Spearman rank correlation coefficient}
The SCC calculation for 2D binary systems in our quiescent measurements is performed using the Fisher-Yates shuffle algorithm, as outlined in section H. The SCC over a time interval $t$ is calculated by considering several configurations or images that are separated by time $t$. This data is collated to form a large dataset. From this pool, several smaller datasets are created using the Fisher-Yates shuffle algorithm. The average value and error bars are then calculated from these smaller sets.\\

The SCC calculation for sheared systems is done using two configurations separated by a strain interval $\Delta \gamma$ in the steady state. The SCC is computed for these configurations and then averaged over several instances of $\Delta \gamma$.\\

\subsection{Coarse-graining of the structural order parameter}

The SOP presented in this manuscript is obtained from the coarse-grained values of the depth of the caging potential. The coarse-grained potential $\Phi_{CG}$ is given by the following expressions
\begin{equation}
\Phi^{i}_{CG} = \frac{\sum_{j} \Phi^{j} f(r_{ij},L)+\Phi^{i}}{\sum_{j} f(r_{ij},L)+1},
\label{eq6}
\end{equation}
where $f(r_{ij},L)$ is the switch function with a cutoff $L$ \cite{Parrinello17_1, Parrinello17_2}. It ensures a value of $1$ for $r_{ij}<<L$, $0$ for $r_{ij}>>L$. The form of this function is given by:
\begin{equation}
f(r_{ij},L) = \frac{1-(r_{ij}/L)^N}{1-(r_{ij}/L)^M},
\end{equation}
where $N=6$, $M=12$, and the cutoff $L = 2$ in units of $\sigma_l$ which is the diameter of the large particles. The value of $L=2$, unless specified explicitly. We adopt this procedure to coarse-grain the local caging potential, and the resulting distributions of SOP are shown in the manuscript.

\section*{Acknowledgements}
The authors are grateful to Smarajit Karmakar, Chandan Dasgupta, Andrea Liu, and Srikanth Sastry for helpful discussions. V.C. acknowledges funding from IISER Pune as a start-up grant and DST-SERB India under grant no. SRG/2019/001922. S.B. acknowledges DST-SERB for POWER fellowship (SERB, Grant No. SPF/2021/000112 ). R.S. and M.S. acknowledge CSIR, India, for PhD fellowships and IISER Pune for PARAM Brahma HPC facility.

%\bibliography{Shear-ref}

%

\end{document}